\documentclass[%
  reprint,
superscriptaddress,
showpacs,
nofootinbib,
nobibnotes,
 amsmath,amssymb,
 aps,
pra,
]{revtex4-1}

\usepackage{graphicx}
\usepackage{dcolumn}

\usepackage{dsfont}

\usepackage{color}
\usepackage[english]{babel}
\usepackage[T1]{fontenc}
\usepackage{todonotes}

\newcommand{\ad}    [1] { \hat{a} ^\dagger _{#1 } }
\renewcommand{\a}   [1] { \hat{a}   _{#1 } }

\newcommand{\tr}{ {\rm tr}}

\begin{document}

\title{The BBGKY hierarchy for ultracold bosonic systems: II. Applications}

\author{Sven Kr\"onke}
	\email{skroenke@physnet.uni-hamburg.de}
	\affiliation{Zentrum f\"ur Optische Quantentechnologien, Universit\"at
Hamburg, Luruper Chaussee 149, 22761 Hamburg, Germany}
	\affiliation{The Hamburg Centre for Ultrafast Imaging, Universit\"at 
Hamburg, Luruper Chaussee 149, 22761 Hamburg, Germany}
\author{Peter Schmelcher}
	\email{pschmelc@physnet.uni-hamburg.de}
	\affiliation{Zentrum f\"ur Optische Quantentechnologien, Universit\"at 
Hamburg, Luruper Chaussee 149, 22761 Hamburg, Germany}
	\affiliation{The Hamburg Centre for Ultrafast Imaging, Universit\"at 
Hamburg, Luruper Chaussee 149, 22761 Hamburg, Germany}

\date{\today}

\begin{abstract}
In this work, which is based on our previously derived theoretical framework
\cite{BBGKY_1},
we apply the truncated Born-Bogoliubov-Green-Kirkwood-Yvon (BBGKY) hierarchy for ultracold
bosonic systems with a fixed number of particles to two out-of-equilibrium 
scenarios, namely tunneling in a double-well potential and
an interaction quench in a harmonic trap. The efficient formulation of the
theory provided in \cite{BBGKY_1} allows for going to large truncation orders
such that the impact of the truncation order on the accuracy of the 
results can be systematically explored. While the short-time dynamics is
 found to be excellently described with controllable accuracy, significant
 deviations occur on a longer time-scale for 
a sufficiently strong interaction quench or the tunneling
scenario. Theses deviations are
accompanied by exponential-like instabilities leading to unphysical
results.  The phenomenology
of these instabilities is investigated in detail and we show that
the minimal-invasive correction algorithm of the equation of motion as proposed in
 \cite{BBGKY_1} can indeed stabilize the BBGKY hierarchy truncated 
 at the second order. 
\end{abstract}

\maketitle
\section{Introduction}\label{sec:intro}

Understanding the impact of correlations on many-body quantum systems constitutes 
a major prerequisite for designing and controlling quantum matter. In these regards,
ultracold atoms provide a highly flexible platform due to their isolatedness 
from an environment
and their controllability, which allows for realizing a great variety of physical situations
(variable trap geometries and interactions)
and thereby for studying many-body effects in a systematic, unprecedented manner \cite{Pethick_Smith2008,many_body_physics_bloch,atoms_lattice_quantum_sim__book2012}.
At the same time, this flexibility constitutes a challenge for theoretical 
approaches due to the wide range of particle numbers and interaction strengths involved.

While there has been tremendous progress in developing highly advanced wavefunction
propagation method such as the time-dependent tensor-network approaches \cite{DMRG_at_age_of_MPS_Schollwoeck2010,out_of_equil_dynamics_with_MPS_Carr_NJP_2012}
or the Multi-Configuration Time-Dependent Hartree method for Bosons (MCTDHB) \cite{MCTDHB_PRA08},
describing an ultracold gas of thousands or even hundred-thousands of atoms with the 
inherent correlations taken accurately into account is not within the immediate reach of methods aiming at the full
many-body wavefunction. In such situations, an effective description of relevant subsystems of the complete
many-body system appears to be much more appropriate. Here, we consider such a
description of reduced density operators corresponding to few-particle subsystems of an 
ultracold bosonic
quantum gas by means of the appropriately truncated Born-Bogoliubov-Green-Kirkwood-Yvon (BBGKY) hierarchy of equations of motion
in its quantum version
\cite{bogoliubov65,born_green47,kirkwood46,yvon57,qua_kin_theo_bonitz}.

While there are numerous theoretical works on the BBGKY hierarchy and 
its truncation (see the references in \cite{BBGKY_1} as well as e.g.\ \cite{qua_kin_theo_bonitz}
for an overview), the literature on the accuracy and stability
of the truncated BBGKY equations of motion (EOM) in dependence on the truncation order by 
explicit numerical simulations
is - to the best of our knowledge - limited \cite{akbari_challenges_2012,jeffcoat_n-representability-driven_2014,schuck14,schuck17,lackner_propagating_2015,lackner_high-harmonic_2017,elliott_density-matrix_2016}.
Actually, most studies deal with fermions (for bosons, see \cite{kira_excitation_2014,kira_coherent_2015,kira_hyperbolic_2015}
 as well the BBGKY-related approaches 
\cite{kohler_microscopic_2002,kohler_microscopic_2003,vardi_bose-einstein_2001,anglin_dynamics_2001,tikhonenkov_quantum_2007,trimborn_decay_2011,witthaut_beyond_2011,kordas_dissipative_2015-1})
and are based on the truncation of the BBGKY hierarchy after the second order.

For this reason, the purpose of this work is to comprehensively study the accuracy of
BBGKY simulations of the non-equilibrium dynamics of finite ultracold bosonic
ensembles when systematically increasing the truncation order in the framework
of a cluster-expansion based closure approximation. While the immediately preceding work \cite{BBGKY_1} covers 
the underlying
theory and its efficient formulation, which allows for going to large truncation orders, we are 
here solely concerned with evaluating its accuracy and stability by numerical applications in two physically very transparent
scenarios. The results for various truncation orders $\bar o$ 
are compared to  the MCTDHB simulations for the full $N$-particle wavefunction.

The first scenario is concerned with the tunneling dynamics
of bosonic atoms in a double-well potential. Treating
the system in the lowest-band tight-binding approximation allows us to 
go to large truncation orders without the need of dynamically optimizing the 
single-particle basis via the corresponding MCTDHB EOM \cite{MCTDHB_PRA08}. 
Thereby, we probe
solely effects stemming from the truncation of the BBGKY hierarchy. Here, we
find the short-time dynamics to be excellently described by the BBGKY hierarchy
and the accuracy to increase monotonously with increasing truncation order. 
For longer times, strong deviations are observed, which are linked to $N$-particle
correlations becoming dominant as well as exponential instabilities of the 
truncated BBGKY EOM resulting in unphysical solutions. 
The phenomenology of these instabilities is analyzed in detail and we show that the
minimal-invasive correction scheme for the BBGKY EOM (truncated at
order $\bar o=2$) \cite{BBGKY_1} can stabilize these EOM indeed.

In the second scenario,
we consider a harmonically trapped bosonic ensemble subjected to an interaction
quench from the non-interacting regime to finite inter-particle repulsion.
Here, we use the fully coupled system of the BBGKY EOM (6) 
of \cite{BBGKY_1}
and the MCTDHB EOM (5) 
of \cite{BBGKY_1} for the time-dependent, variationally optimized single-particle basis.
For low excitation energies, we find the system to be highly accurately described by the
truncated BBGKY approach. For higher excitation energies, however, exponential instabilities
again occur. Also in this case, we can stabilize the BBGKY EOM truncated at $\bar o=2$ by
our EOM correction scheme and obtain reasonably accurate results for longer times.

This work is organized as follows. While the main text is focused
on the applications, namely the tunneling scenario in Section \ref{sec:BH_dimer} and the
interaction-quench scenario in Section \ref{sec:breath}, technical comments on the numerical integration
of the truncated BBGKY EOM are provided in Appendix \ref{app:integration}. 
Finally, we conclude in Section \ref{sec:concl}.
\section{Tunneling dynamics in a Bose-Hubbard dimer}\label{sec:BH_dimer}

In this scenario, we assume that $N$ bosonic atoms are loaded into an effectively
one-dimensional double-well potential. Preparing the system in an initial state featuring a
particle-number imbalance with a left and the right well allows for studying
the tunneling dynamics of such a many-body system, which has been subject of 
numerous studies covering both mean-field \cite{raghavan_coherent_1999,smerzi_quantum_1997} and many-body calculations 
taking correlations into account \cite{milburn_quantum_1997,exact_quantum_dynamics_Cederbaum_PRL2009,sakmann_quantum_2010,univ_frag_dw,gertjerenken_beyond-mean-field_2013}. Effects unraveled in such a realization
of a bosonic Josephson junction cover macroscopic tunneling and self-trapping \cite{raghavan_coherent_1999,smerzi_quantum_1997,direct_obs_boson_JJ}
as well a decay of tunneling oscillations due to the dephasing of populated
many-body eigenstates of the post-quench Hamiltonian \cite{milburn_quantum_1997,exact_quantum_dynamics_Cederbaum_PRL2009,sakmann_quantum_2010,univ_frag_dw,gertjerenken_beyond-mean-field_2013}.

For sufficiently deep wells, the microscopic many-body Hamiltonian of this
system can be well
approximated by a two-site Bose-Hubbard Hamiltonian within the 
lowest-band tight-binding approximation
\begin{align}
\label{eq:BH_hamilt}
 \hat H = &-J(\ad{L}\a{R}+\ad{R}\a{L})\\\nonumber
 &+\frac{U}{2}\big[\hat n_L(\hat n_L-1)
 +\hat n_R(\hat n_R-1)
 \big]
\end{align}
where $\a{L/R}$ annihilates a boson in the lowest-band Wannier state localized in
the left/right well and $\hat n_{i}\equiv\ad{i}\a{i}$ denotes the corresponding 
occupation-number operator of the site $i\in\{L,R\}$. The first
term in \eqref{eq:BH_hamilt} describes tunneling between the two wells
weighted with the hopping amplitude $J>0$. The second term refers to
on-site interaction of strength $U$ and stems from the 
short-range van-der-Waals interaction between the atoms. For convenience,
we take the hopping amplitude as our energy scale and state times
in units of $1/J$.

The Bose-Hubbard dimer features an almost trivial computational complexity since
the full many-body wavefunction depends only on $C^N_2=N+1$ complex-valued 
coefficients such that the corresponding time-dependent Schr\"odinger equation can be 
numerically exactly solved for very large atom numbers. So there is no
need for an alternative computational approach here. On the other hand,
this system can serve as a good playground for analyzing the properties of the
truncated BBGKY approach because (i) the corresponding numerically exact solution
is available and (ii) we can easily represent RDMs of large order without using
a dynamically optimized truncated single-particle basis. This allows for 
systematically investigating the accuracy of our results solely in dependence on
the truncation order $\bar o$.

In the following, 
we consider the initial state $|\Psi_0\rangle=|N,0\rangle$ with all atoms located in the
left well and focus on the tunneling regime by setting the dimensionless interaction parameter
$\Lambda=U(N-1)/(2J)$ to $0.1$, i.e.\ well below the critical value $\Lambda_{\rm crit}=2$
for self-trapping \cite{raghavan_coherent_1999,smerzi_quantum_1997}. In the weak interaction
regime $\Lambda\ll1$, beyond mean-field effects such as
the aforementioned collapse of tunneling oscillation \cite{milburn_quantum_1997} and the universal formation of 
a two-fold fragmented condensate out of a single condensate \cite{univ_frag_dw}
are expected to
play a significant role after the time-scale $t_{\rm mf}\approx\sqrt{2N+1}/(J\Lambda)$,
the so-called quantum break time
\cite{gertjerenken_beyond-mean-field_2013}.

Most of the following calculations deal with $N=10$ atoms such that $t_{\rm mf} \approx 46/J$.
For comparison, we also increase $N$ to $100$ atoms while keeping
$\Lambda$ constant, which results in a longer quantum break time of $t_{\rm mf}  \approx 142/J$.
We analyze the accuracy of the truncated BBGKY hierarchy approach in three steps. First,
we inspect the particle-number imbalance, a highly-integrated quantity 
characterizing the tunneling dynamics, second turn to the eigenvalues of the lowest-order RDMs, which constitute a highly
sensitive measure for correlations, and third compare the whole
lowest order RDMs to the corresponding exact results. For a deeper interpretation
of these findings, we finally analyze the exact results for the
whole $N$-particle wavefunction as well as for the corresponding $o$-particle
correlations. Finally, we investigate the performance of the
correction strategies outlined in Section V B 
of \cite{BBGKY_1}.

\subsection{Particle-number imbalance}

\begin{figure}[t]
\centering
 \includegraphics[width=0.495\textwidth]{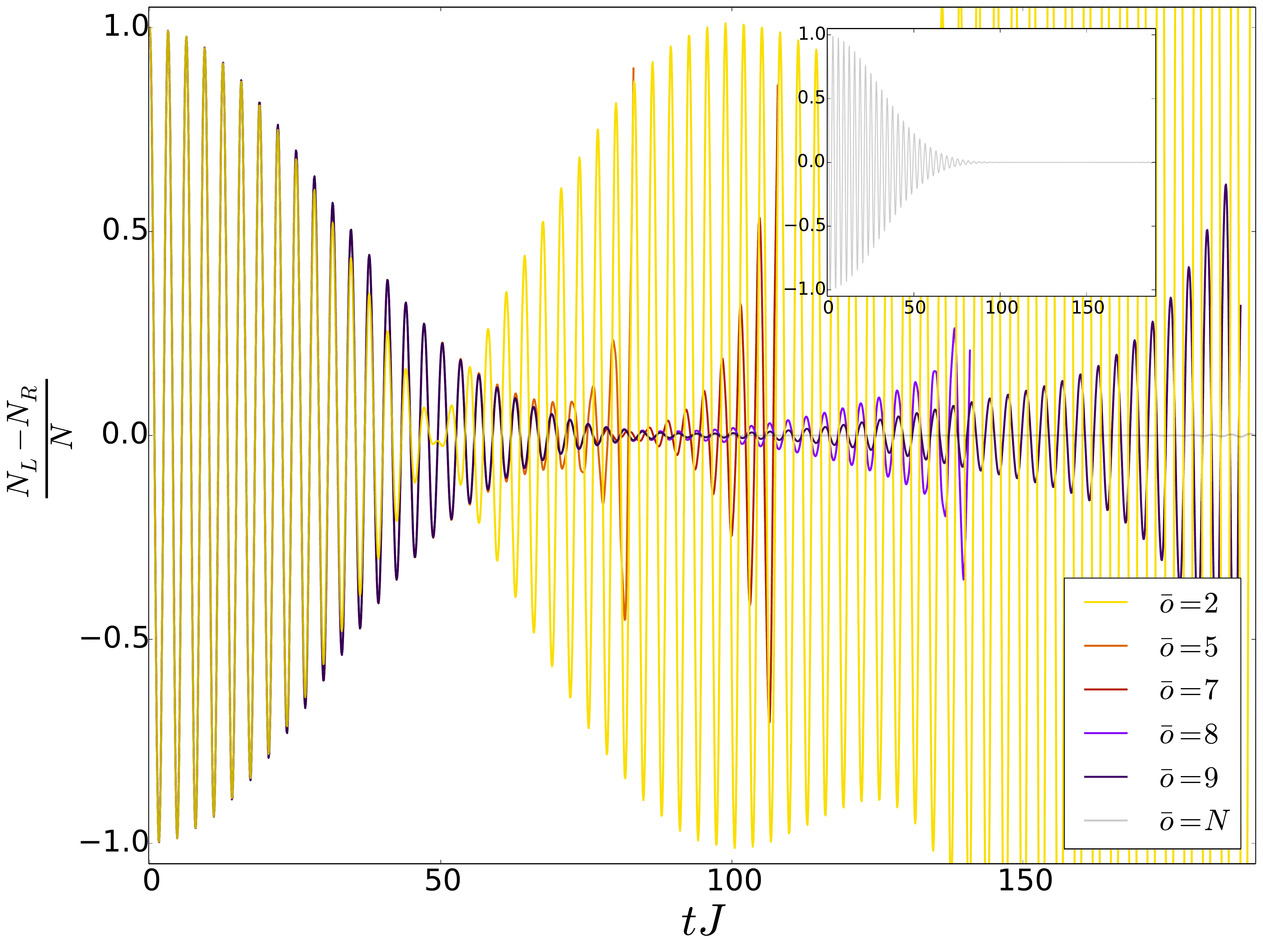}
 \caption{(color online) Time evolution of the particle-number imbalance $(N_L-N_R)/N$
 with $N_i\equiv\langle\hat n_i\rangle$, $i\in\{L,R\}$ for various
 truncation orders $\bar o$. Inset: numerically exact solution of the many-body Schr\"odinger
 equation. Parameters: $N=10$ atoms located initially in the left
 well, dimensionless interaction parameter $\Lambda=0.1$.}
 \label{fig:BH_imbalance}
 \end{figure}

In order to study the tunneling dynamics, the imbalance of the particle 
numbers between the left and right well, $[\langle\hat n_L\rangle-\langle\hat n_R\rangle]/N$,
is depicted in Figure \ref{fig:BH_imbalance} for $N=10$ atoms.
Focusing first on the inset, which shows the numerically exact results,
we see the expected collapse of tunneling oscillations due to a dephasing
of the populated post-quench Hamiltonian eigenstates. Indeed, this collapse
happens on the time-scale $t_{\rm mf} \approx 46/J$, while a 
corresponding Gross-Pitaevskii
mean-field simulation would reveal undamped tunneling oscillations (not shown,
see e.g.\ \cite{gertjerenken_beyond-mean-field_2013}). After $t\sim 200/J$, a
revival of the tunneling oscillations emerges in the numerically exact
calculation (not shown). 

Turning now to the truncated BBGKY approach, we see that all truncation orders
$\bar o\geq2$ give good results for the first $\sim 8$ tunneling oscillations.
Thereafter, the $\bar o=2$ curves departs from both the exact
and the higher truncation-order results, and features a premature maximal
suppression of tunneling oscillations at $t\sim50/J$. In
the subsequent premature revival of tunneling oscillations unphysical 
values $|\langle\hat n_L\rangle-\langle\hat n_R\rangle|/N>1$ are reached
at about $t=100/J$, indicating a lack of $1$-RDM representability.

These findings suggest that higher-order correlations than $\hat c_2$ (see Section IV B 3 
of \cite{BBGKY_1} for the definition) play a 
significant role. Increasing the truncation order $\bar o$ stepwise up to
the maximally reasonable order $\bar o=N-1=9$,
we clearly see that the accuracy of our results improves systematically.
The larger $\bar o$ is, the more accurate is the collapse of the tunneling
oscillations described. However, all non-trivial truncations $\bar o<N$
predict a premature revival of the tunneling oscillations, which goes
hand and in hand with a maximal suppression of the tunneling-oscillation
amplitude to small but noticeable values (while the exact results
do not feature noticeable oscillations at the corresponding times).
We note that for $2<\bar o<10$ the simulations suffer from drastic instabilities
of the EOM,
being discussed in the subsequent Section, such that
we had to stop them after a certain time. This is why the corresponding
curves in Figure \ref{fig:BH_imbalance} are not provided for the
whole range of depicted times.

\subsection{Natural populations} 
 \begin{figure}[t]
   \includegraphics[width=0.495\textwidth]{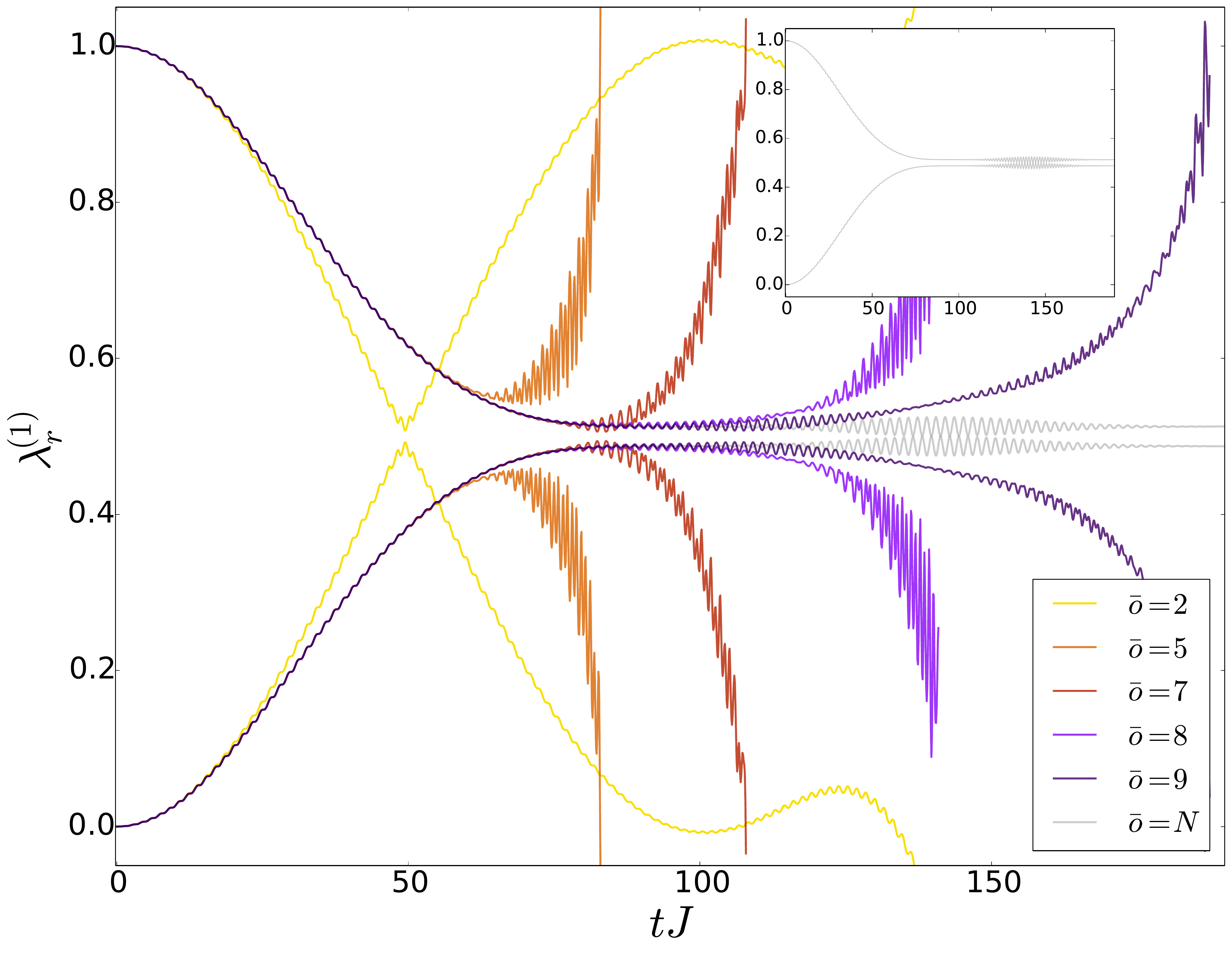}
   \caption{(color online) Natural populations of the $1$-RDM for various
   truncation orders $\bar o$. Inset: numerically exact solution of the many-body Schr\"odinger
 equation. Parameters: same as in Figure \ref{fig:BH_imbalance}.}
   \label{fig:BH_npop1}
\end{figure}
 \begin{figure}[t]
   \includegraphics[width=0.495\textwidth]{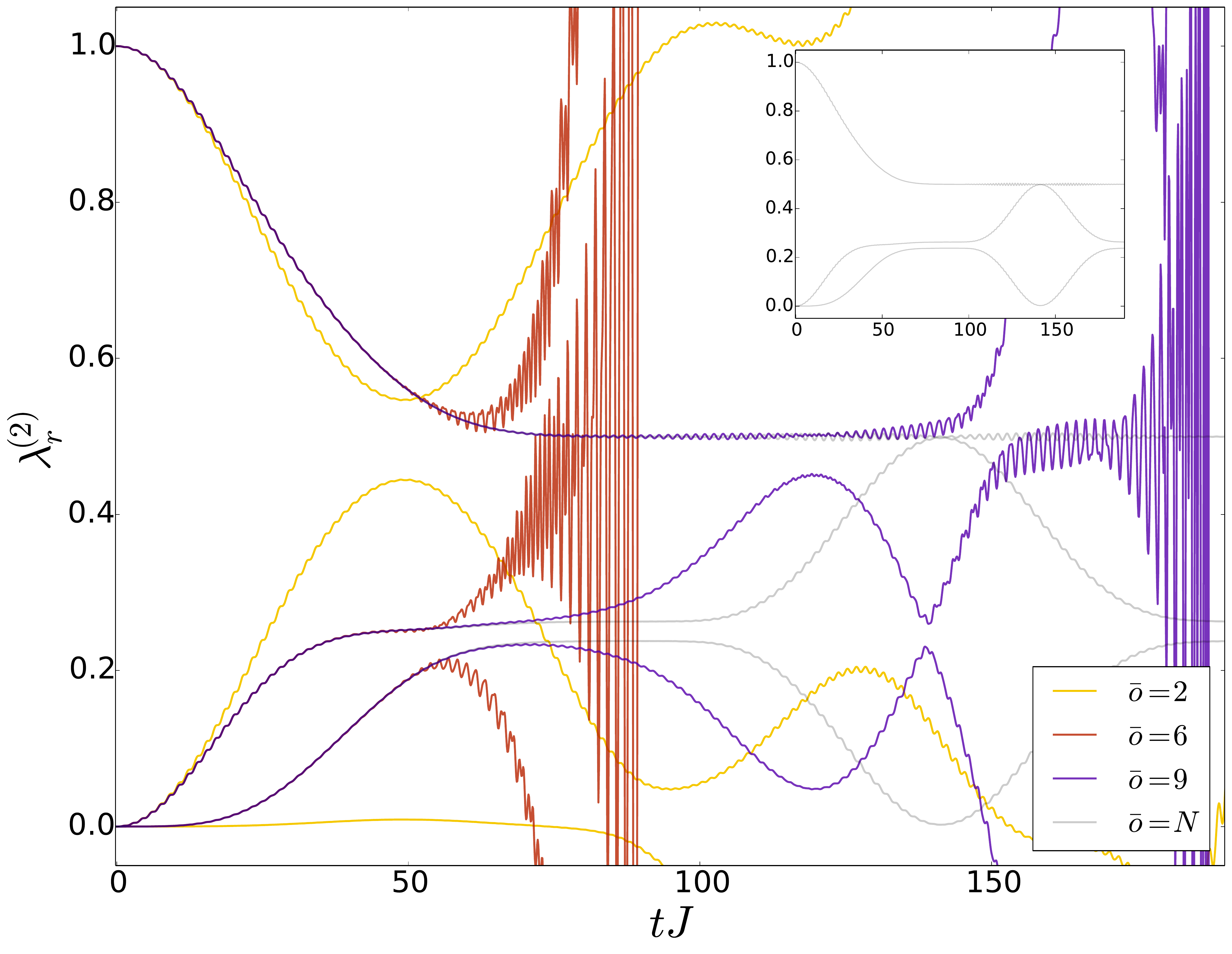}
   \caption{(color online) Natural populations of the $2$-RDM for various
   truncation orders $\bar o$. Inset: numerically exact solution of the many-body Schr\"odinger
 equation. Parameters: same as in Figure \ref{fig:BH_imbalance}.}
   \label{fig:BH_npop2}
\end{figure}

Next we analyze the eigenvalues $\lambda^{(o)}_i$  of the $o$-RDM, called natural populations\footnote{The terms 'natural population' and
'natural orbital' have originally been introduced for the 1-RDM only \cite{loewdin_norb55}
but are employed for all orders in this work.} (NPs) in the following and 
start with the NPs of the $1$-RDM NPs in Figure 
\ref{fig:BH_npop1}, which can diagnose beyond mean-field behavior. 
The numerically exact results (see corresponding inset)
reveal dynamical quantum depletion leading to 
a two-fold fragmented condensate for $t\gtrsim 80/J$ with almost equal population
of the corresponding NOs, $\lambda_1^{(1)}\approx0.5\approx\lambda_2^{(1)}$ (see also e.g.\
\cite{univ_frag_dw}). Strikingly fast oscillations in these NPs emerge and decay
around $t\sim140/J$, which we can connect to the periodical emergence and decay
of a NOON state of the total system (see below).

The corresponding results of the truncated BBGKY approach feature a similar 
dependence on the truncation order $\bar o$ as the particle-number imbalance does.
While the $\bar o=2$ prediction starts to deviate noticeably from the exact results already
for $t\gtrsim25/J$, we obtain trustworthy results for a longer time, the larger
$\bar o$ is chosen. In particular, the truncated BBGKY approach can accurately
determine the achieved mean degree of fragmentation 
(see $\bar o=8,9$ results at $t\sim100/J$). Even for the largest truncation order $\bar o=9$, however,
the truncated BBGKY simulations 
predict a premature and very fast 
revival of condensation (i.e. $\lambda_1^{(1)}\approx 1$), while 
this process starts only after $t\sim 200/J$ in the exact calculation and happens
more slowly (not shown).
Most importantly, this unphysical fast re-condensation overshoots the
range of valid NPs such that the $1$-RDM ceases to be positive semi-definite, indicating
an exponential-like instability of the EOM. 

While we have so far only studied the prediction of the truncated BBGKY approach
for one-particle properties, 
we now inspect the NPs of the $2$-RDM in Figure
\ref{fig:BH_npop2}, also called
natural geminal populations \cite{rdm_Sakmann_PRA2008}. 
The exact dynamics (see the inset) features two important aspects, which we have
also observed for the NPs of higher-order RDMs (not shown). (i)
The dominant NP $\lambda_1^{(2)}$ first loses weight in favor for 
the other NPs. (ii) At about $t\sim140/J$, all NPs are suppressed except for
$\lambda_1^{(2)}\approx 0.5\approx\lambda_2^{(2)}$. Having observed
the latter feature for the NPs of all orders $o\in\{1,...,9\}$, we may conclude that 
in this stage of the dynamics a subsystem of $o$ particles occupies approximately
only two $o$-particle states with almost equal probabilities. As we will see below,
this finding is caused by the periodical emergence and decay
of a NOON state of the total system which is discussed below.

Turning now to the predictions of the truncated BBGKY approach, we 
see again a systematic improvement of accuracy with increasing truncation
order $\bar o$. The maximal time for which the highest truncation order 
$\bar o=9$ gives reliable results, however, has reduced from $t\sim110/J$ 
for the $1$-RDM NPs (see Figure \ref{fig:BH_npop1}) to 
$t\sim70/J$ for the $2$-RDM NPs (see Figure \ref{fig:BH_npop2}). Thereafter,
the largest NP $\lambda_1^{(2)}$ is well described until $t\sim130/J$,
while the other two NPs already show strong deviations: it seems that the emergence of 
the feature (ii) discussed above happens premature, name at about $t\sim120/J$.
Furthermore, we also witness the exponential-like instabilities leading to 
$2$-RDM NPs outside the interval $[0,1]$.

\begin{figure*}[t]
 \includegraphics[width=0.495\textwidth]{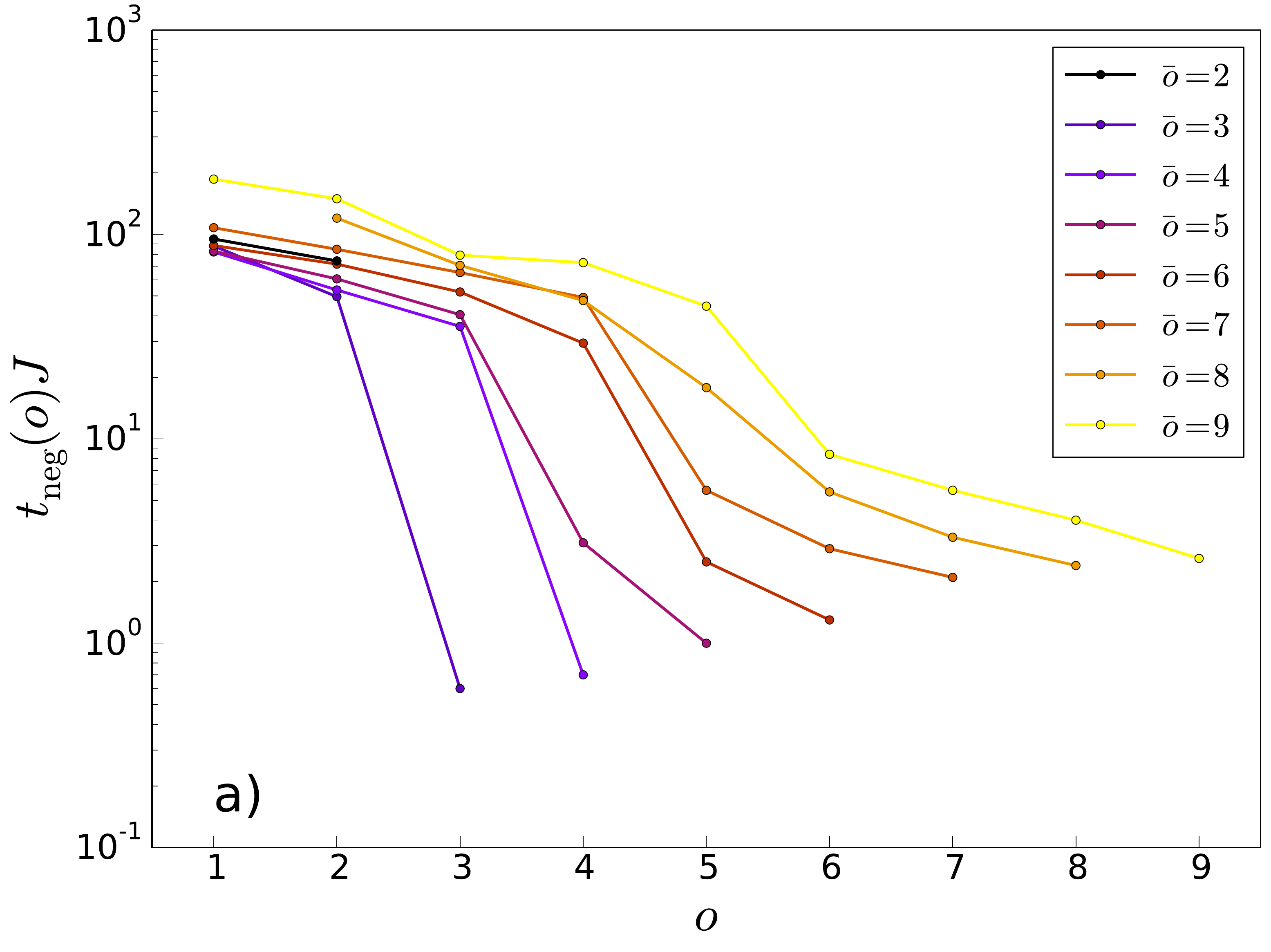}
 \includegraphics[width=0.495\textwidth]{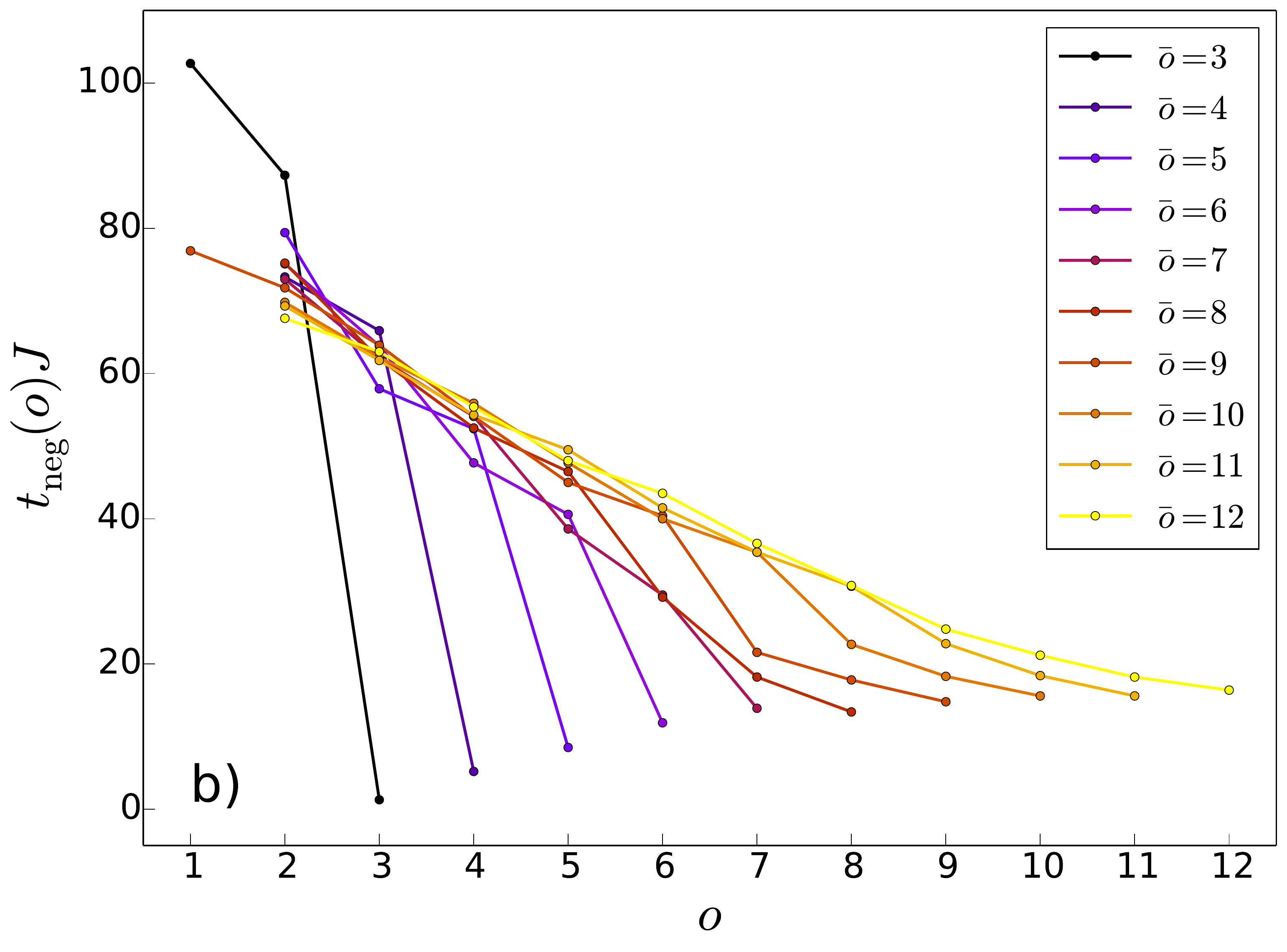}
 \caption{(color online) First time $t_{\rm neg}(o)$ when the lowest $o$-RDM NP is smaller than $\epsilon=-10^{-10}$
 in dependence on $o$ for various truncation orders $\bar o$.
 a) Same parameters as in Figure \ref{fig:BH_imbalance}. b) Same as a) but for the atom number $N$ increased to $100$
 while keeping the interaction parameter $\Lambda=0.1$ constant. The results for $\bar o=2$ are not plotted in b)
 and read $t_{\rm neg}(1)\approx216/J$ as well as  $t_{\rm neg}(2)\approx168/J$. }
 \label{fig:BH_neg_t}
\end{figure*}

In order to analyze how this unphysical behavior emerges, we depict the first
time $t_{\rm neg}(o)$ when the lowest $o$-RDM NP $\lambda^{(o)}_{o+1}$ 
is smaller than the threshold $\epsilon=-10^{-10}$ for various
$o$ and different truncation orders $\bar o$ in Figure \ref{fig:BH_neg_t} a).
For fixed truncation order $\bar o$, $t_{\rm neg}(o)$ decreases 
with increasing order $o$. This means that the representability defect
of $\hat\rho_o$ lacking positive semi-definiteness starts at the
truncation order $o=\bar o$ and propagates then successively to
lower orders due to coupling via the collision integral. For most
orders $o$, we moreover find that $t_{\rm neg}(o)$ increases with
increasing truncation order $\bar o$, which fits to the above findings regarding
enhanced accuracy for larger $\bar o$ (exceptions occur at order $o=1,2$  in particular
for $\bar o=2$). 

Increasing the number of atoms to $N=100$ while keeping the dimensionless interaction parameter
$\Lambda=0.1$ constant, we again find a monotonous decrease of $t_{\rm neg}(o)$ with increasing $o$
for fixed truncation order $\bar o$ (see Figure \ref{fig:BH_neg_t} b)). This confirms
the above finding that the lack of positivity successively propagates from higher to lower orders.
In contrast to the $N=10$ case, we only find an enhancement of $t_{\rm neg}(o)$ with increasing
$\bar o$ for orders $o\geq6$. In particular, we see that the largest truncation order considered,
$\bar o=12$, features the smallest $t_{\rm neg}(o=2)$. It is 
quite possible that the an ``enhancement'' of non-linearity with increasing truncation
order $\bar o$ (note that the applied closure approximation, cf.\ Section IV B 3 
of \cite{BBGKY_1}, is a polynomial of degree $(\bar o+1)$ 
in $\hat\rho_1$ and of degree $\lfloor(\bar o+1) /o\rfloor$ in the cluster $\hat c_o$) is the
reason why the BBGKY EOM are more prone to these instabilities for larger $\bar o$.

Having compared so far only certain aspects of $o$-particle properties, 
we finally aim at comparing the prediction of the truncated BBGKY approach for 
the whole $o$-RDM to the exact results.

\subsection{Reduced density operators}

\begin{figure*}[t]
 \includegraphics[width=0.495\textwidth]{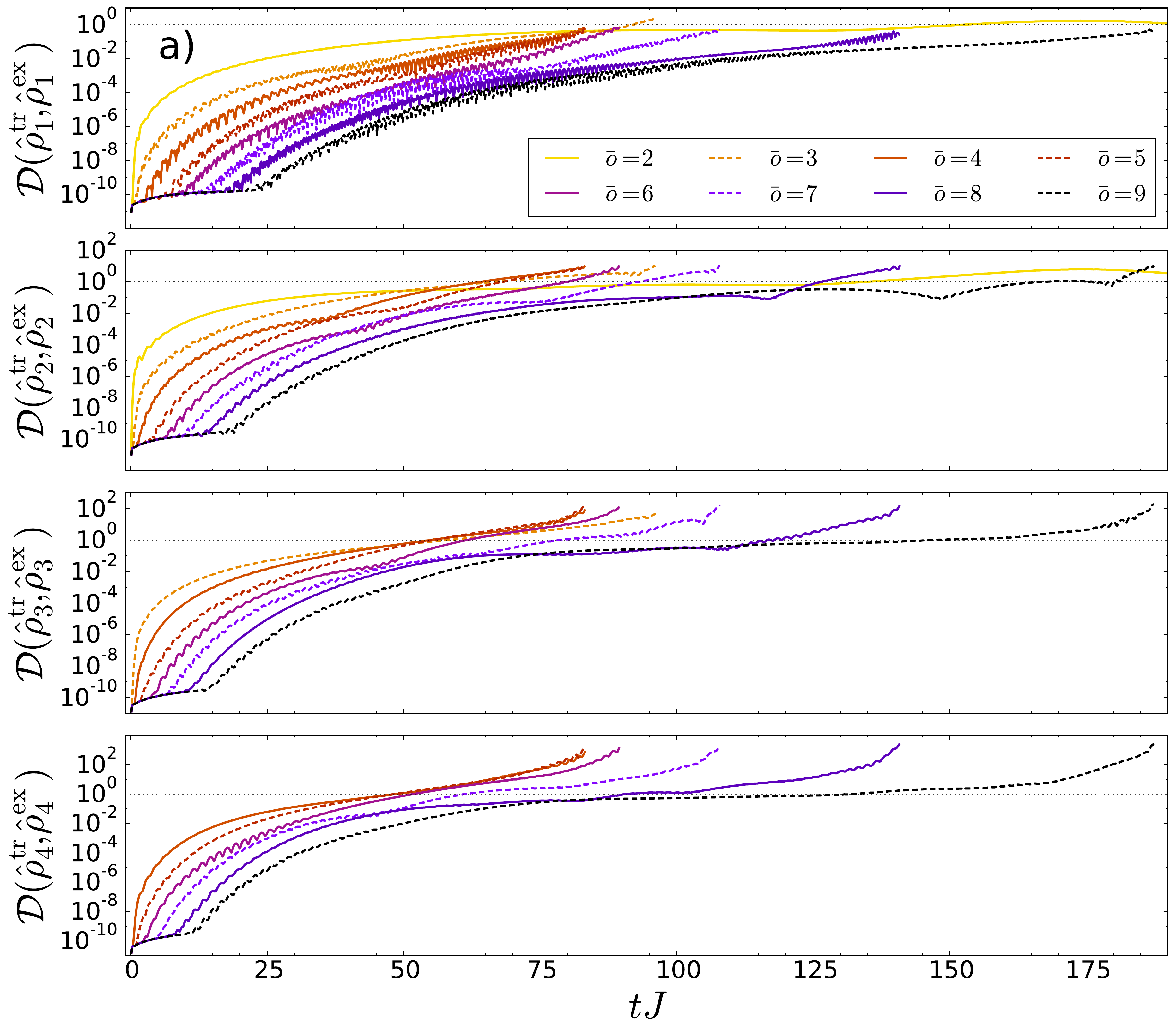}
  \includegraphics[width=0.495\textwidth]{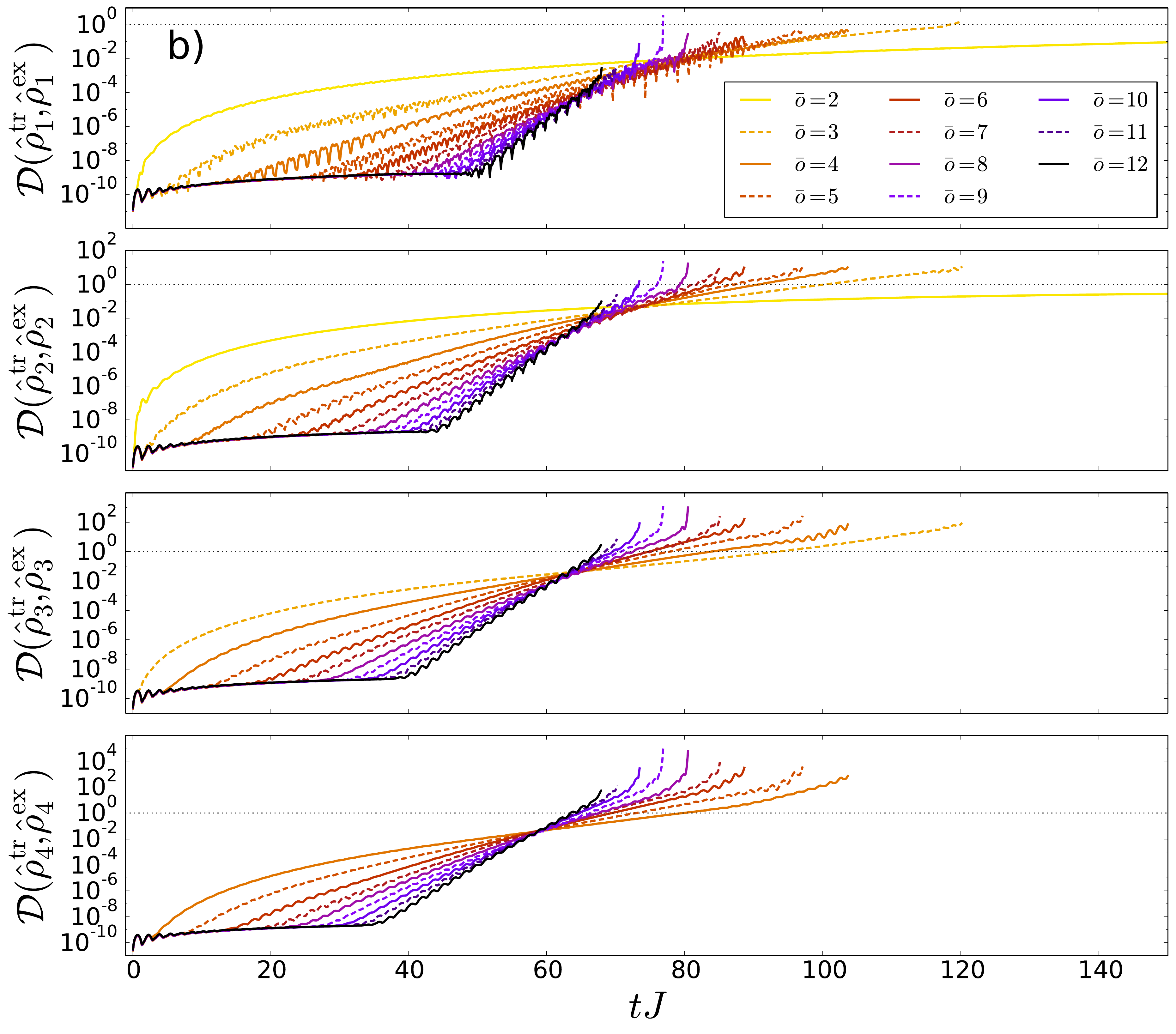}
 \caption{(color online) Time evolution of the trace distance $\mathcal{D}(\hat\rho_o^{\rm tr},\hat\rho_o^{\rm ex})$
 between the exact result and the truncated BBGKY prediction for the $o$-RDM ($o=1,...,4$) and various truncation
 orders $\bar o$. The dotted horizontal lines at unity ordinate value indicate the upper bound for the trace distance
 between two density operators (see main text).
 a) Same parameters as in Figure \ref{fig:BH_imbalance}. b) Same as a) but for the atom number $N$ increased to $100$
 while keeping the interaction parameter $\Lambda=0.1$ constant.}
 \label{fig:BH_rdm_comp}
\end{figure*}

For this purpose, we 
take the trace distance
$\mathcal{D}(\hat\rho_o^{\rm tr},\hat\rho_o^{\rm ex})\equiv||\hat\rho_o^{\rm tr}-\hat\rho_o^{\rm ex}||_1/2$
\cite{nielsen_chuang_book} as a measure for deviations between the
truncated BBGKY prediction for the $o$-RDM denoted by $\hat\rho_o^{\rm tr}$ and the
numerically exact result $\hat\rho_o^{\rm ex}$. Here, $||\cdot||_1$ refers to the trace-class norm (also 
called Schatten-1 norm) being defined as $||\hat A||_1\equiv\tr(\sqrt{\hat A^\dagger\hat A})$
for any trace-class operator $\hat A$. For hermitian operators $\hat A$, $||\hat A||_1$ 
equals the sum of absolute values of $\hat A$'s eigenvalues. 
By means of $\mathcal{D}(\hat\rho_o^{\rm tr},\hat\rho_o^{\rm ex})$,
one can estimate the maximal difference for the expectation values of any (trace-class)
$o$-body observable $\hat A_o$ as follows $|\tr(\hat A_o\hat\rho_o^{\rm tr})-\tr(\hat A_o\hat\rho_o^{\rm ex})|
\leq 2||\hat A_o||_1\,\mathcal{D}(\hat\rho_o^{\rm tr},\hat\rho_o^{\rm ex})$, whose
proof is reviewed in Appendix \ref{app:exp_value_bound}. Moreover,
given that its arguments are density operators (i.e. hermitian, positive semi-definite
and trace one), the trace-distance is bounded by
$\mathcal{D}(\hat\rho_o^{\rm tr},\hat\rho_o^{\rm ex})\in[0,1]$ and
can be interpreted as the
probability that these two quantum states can be distinguished by the outcome of a single measurement 
\cite{nielsen_chuang_book}. 

In Figure \ref{fig:BH_rdm_comp}, we depict $\mathcal{D}(\hat\rho_o^{\rm tr},\hat\rho_o^{\rm ex})$
for the orders $o=1,...,4$ and various truncation orders $\bar o$, where subfigures a) and b) refer
to the $N=10$ and $N=100$ case with the same interaction parameter 
$\Lambda=0.1$, respectively. For fixed truncation order $\bar o$, we clearly see that the accuracy of the 
truncated BBGKY prediction for the $o$-RDM decreases with increasing order $o$. Up to a
certain time, which depends on the order $o$, we moreover find $\mathcal{D}(\hat\rho_o^{\rm tr},\hat\rho_o^{\rm ex})$
to decrease with increasing truncation order $\bar o$. 

The instabilities of the truncated BBGKY EOM manifest themselves in the trace distant exceeding
its upper bound $\mathcal{D}(\hat\rho_o^{\rm tr},\hat\rho_o^{\rm ex})\leq 1$ for density operators,
implying that $\hat\rho_o^{\rm tr}$ lacks to have trace one or to be positive semi-definite. Since 
the conservation of the initial RDM trace is ensured by the truncated BBGKY approach, violations
of $\tr(\hat\rho_o^{\rm tr})=1$ can at most occur numerically if the system gets deep into the
exponential-like instability (where we observe the truncated BBGKY EOM to become stiff
such that the integrator has a hard time). Thus, exceeding the upper bound on the trace distance
is connected to a lack of positive semi-definiteness and can be observed to
happen earlier for increasing order $o$ and fixed truncation order $\bar o$.

For the case of $N=100$ atoms (see Figure \ref{fig:BH_rdm_comp} b)), we observe the additional
particularity that in the vicinity of $t\sim63/J$ the accuracy of the truncated BBGKY prediction
for the $o$-RDM does not depend on the truncation order $\bar o$, which happens slightly earlier
for larger $o$. Before this point, a systematic increase of accuracy is observed for increasing 
truncation order $\bar o$. Thereafter, lower truncation orders give (slightly) better results than
higher ones. Furthermore, while in the $N=10$ case one-body properties (such as e.g.\ the particle-number
imbalance) can be described with reasonable accuracy up to $t\sim2\, t_{\rm mf}$ (when the collapse
of tunneling oscillations has already taken place), the instabilities hinders us to obtain
accurate results for $t$ larger than about $0.56\, t_{\rm mf}$ in the case of $N=100$ (at this time,
the tunneling oscillation amplitude is still significant).

\subsection{Many-body state and $o$-particle correlations: exact results}

\begin{figure*}[t]
 \includegraphics[width=0.7\textwidth]{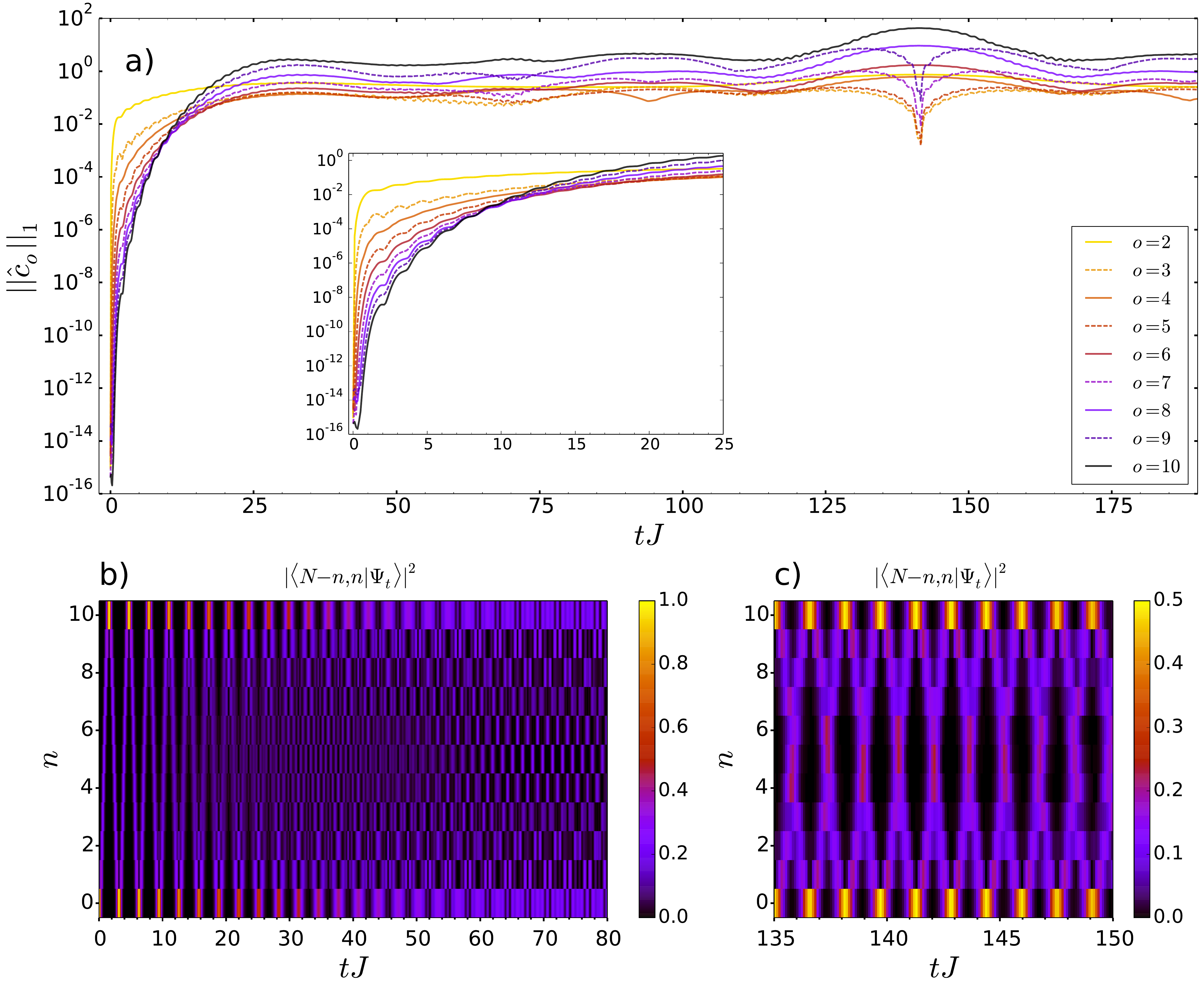}
 \caption{(color online) a) Time evolution of the cluster's trace-class norm $||\hat c_o||_1$
 for all orders $o$, obtained from the numerically exact solution of the
 time-dependent Schr\"odinger equation. Inset: zoom into early time dynamics. b) and c) Probability
 to find $n$ atoms in the right and $(N-n)$ atoms in the left well, $|\langle N-n,n|\Psi_t\rangle|^2$, versus time
 for two characteristic stages of the dynamics. Parameters: same as in Figure \ref{fig:BH_imbalance}.}
  \label{fig:BH_cluster_dynamics}
\end{figure*}

In order to obtain physical insights in the above findings, we finally come back
to the numerically exact results for $N=10$ and measure the strength of $o$-particle correlations (see the
definition given in Section IV B 3 
of \cite{BBGKY_1})
in terms of $||\hat c_o||_1$ in
Figure \ref{fig:BH_cluster_dynamics} a). From the inset, we infer that the correlations
initially build up in a hierarchical manner. First, only two-particle correlations  start to play a role, 
then three-particle correlations and so on. This hierarchy in $||\hat c_o||_1$ holds, however, only until
$t\sim8/J$, when the ordering of the $||\hat c_o||_1$'s with respect to the
order $o$ starts to become reversed. After a certain point, $N$-particle correlations become the most dominant
ones. 
This holds in particular in the vicinity of $t\sim140/J$, where we have observed fast oscillations
in the NPs $\lambda^{(1)}_{1/2}$ and found for all orders $o=1,...,9$ that
the RDMs feature approximately only two finite NPs $\lambda^{(o)}_{1}\approx0.5\approx\lambda^{(o)}_{2}$.
At this stage of the dynamics, all clusters $\hat c_o$ of odd order $o$ are strongly suppressed.

For connecting the above findings regarding $o$-particle correlations to the full many-body state, 
we depict in the Subfigures \ref{fig:BH_cluster_dynamics} b)
and c) the probability $|\langle N-n,n|\Psi_t\rangle|^2$ of finding $n$ atoms in the right and $(N-n)$ atoms in the left well.
For the early dynamics, we witness how the system
becomes delocalized in the Fock space such
that the tunneling oscillations become suppressed (Subfigure \ref{fig:BH_cluster_dynamics} b)). 
At later times, around
$t\sim140/J$, we, however, find the system to periodically oscillate between 
a NOON state $(|N,0\rangle+e^{i\theta}|0,N\rangle)/\sqrt{2}$ (with some phase $\theta\in\mathbb{R}$)
and some broad distribution being approximately symmetric with respect to its maximum 
at about $n=5$ (Subfigure \ref{fig:BH_cluster_dynamics} c)). Due to this approximate symmetry of the 
distribution around $n=5$, the particle-number imbalance approximately equals
$[\langle\hat n_L\rangle-\langle\hat n_R\rangle]/N\approx0.5$, i.e.\ tunneling
oscillations are still suppressed. This approximate symmetry moreover leads to
a doubling of the oscillation frequency compared to the initial tunneling-oscillation frequency,
which is most probably linked to the fast oscillations in  $\lambda^{(1)}_{1/2}$.
Finally, one can analytically show  that 
the $n$-RDM of the above mentioned NOON state reads $\hat\rho_n=(|n,0\rangle\!\langle n,0|+|0,n\rangle\!\langle 0,n|)/2$,
meaning that the state of an $n$-particle subsystem is an incoherent statistical mixture with
all particles residing in the left (right) well with probability $0.5$.
Thereby, we can directly connect the fact that the RDMs of all orders feature approximately only two finite
NPs of approximately equal value to the underlying many-body state. Coming back to the findings for  $||\hat c_o||_1$
of Figure \ref{fig:BH_cluster_dynamics} a), we may conclude that a NOON state 
leads to strong high-order correlations $\hat c_o$ such that 
truncating the BBGKY hierarchy by means of the applied cluster expansion 
cannot be expected to give accurate results.

In summary, we have seen following.  (i) While the truncated BBGKY approach gives highly accurate
results for short times with controllable accuracy via the truncation order $\bar o$,
 the BBGKY approach
shows deviations at longer times. (ii) Exponential-like instabilities, induced
by the non-linear truncation approximation, propagate form high to low orders
and lead to unphysical results at a certain point. 
(iii) $o$-particle correlations arise very fast
in this tunneling scenario and soon cease to be in decreasing order with respect
to $o$. (iv) The system evolves into a NOON state being dominated by $N$-particle
correlations.

There appear to be at least two plausible causes why the BBGKY approach fails at a certain point:
First, the number of terms in the cluster expansion
(23) 
of \cite{BBGKY_1} drastically increases with the order $o$,
which implies that clusters should decay fast for a
controllable approximation. For example, at the largest truncation order
considered above, $\bar o=12$, the truncation approximation $\hat\rho_{13}^{\rm appr}$ 
already involves $100$ terms. Our findings (iii) and (iv), however, 
might indicate that this system is not suitable for 
a truncation based on the $o$-particle correlations defined in
Section IV B 3 
of \cite{BBGKY_1}. Other truncation approximations
might by more suitable.

Second, the exponential-like instabilities, being connected to a 
lack of representability, might be the main cause for the failure of the
BBGKY EOM at longer times. This hypothesis is supported by the 
fast break-down of the BBGKY approach in the $N=100$ case for the truncation
order $\bar o=12$. For this reason, we analyze next the performance of the 
correction strategies outlined in Section V B 
of \cite{BBGKY_1}.

\subsection{Performance of the correction algorithms}

\begin{figure*}[t]
 \includegraphics[width=0.9\textwidth]{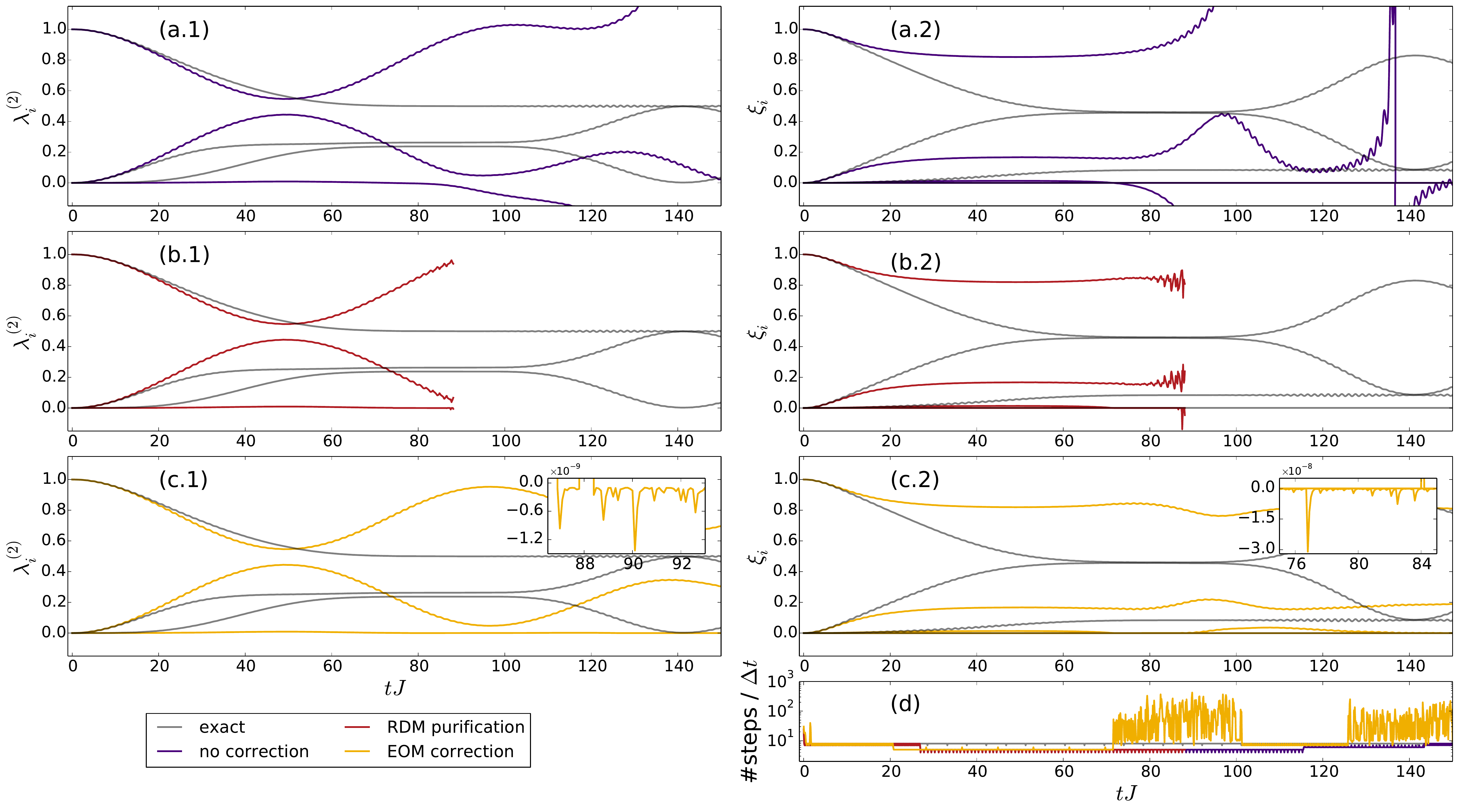}
 \caption{(color online) Comparison of the correction
 strategies outlined in Section V B 
 of \cite{BBGKY_1} (for the BBGKY hierarchy
 truncated at $\bar o=2$). Left column:
 Time evolution of the NPs $\lambda^{(2)}_i$. Right column [except for (d)]:
 Time evolution of the $\hat K_2$ eigenvalues $\xi_i$. First row: Truncated BBGKY
 results without correction versus exact ones. Second row: Truncated BBGKY
 results with iterative minimal-invasive purification of the $2$-RDM after each
 $\Delta t=0.1/J$ (maximal number of iterations: $500$). Third row:
  Truncated BBGKY results with minimal-invasive correction of the $2$-RDM EOM
  (damping rate of negative eigenvalues: $\eta=10J$). Insets of (c.1) and (c.2):
  close-ups showing the imposed exponential damping of negative eigenvalues.
  For both correction strategies, eigenvalues are regarded as negative if they 
  are smaller than the threshold $\epsilon=-10^{-10}$. Subfigure (d): number 
  of ZVODE integrator \cite{zvode} steps per write-out time-step $\Delta t$.
  Parameters: same as in Figure \ref{fig:BH_imbalance}.}
  \label{fig:BH_correction_alg}
\end{figure*}

In the following, we first focus on the correction algorithms applied 
to the BBGKY hierarchy truncated at $\bar o=2$. Thereafter, we comment on
the performance of these algorithms if extended to larger truncation orders
by means of a corresponding ansatz for the correction operator (see Section V B 
of \cite{BBGKY_1}).

Figure \ref{fig:BH_correction_alg} depicts the time evolution of the NPs $\lambda^{(2)}_i$
and the $\hat K_2$ eigenvalues $\xi_i$ (see the definition given in Section V A 
of \cite{BBGKY_1}) for the truncated BBGKY results without correction,
with the iterative minimal invasive purification of the $2$-RDM and the minimal
invasive correction of the $2$-RDM EOM in comparison to the exact results.
Apparently, all cases deviate significantly from the exact results after 
$t\gtrsim17/J$ so that we shall concentrate here solely on the stabilization
performance of the correction algorithms. 

Inspecting first the uncorrected
results [Figures \ref{fig:BH_correction_alg} (a.1), (a.2)], we observe that 
the $K$-condition\footnote{We call this condition $K$- instead of $G$-condition, as commonly done, in order to stress that we operate
with the original, more restrictive definition of the one-particle-one-hole RDM of \cite{garrod_1964}.} (i.e.\ $\hat K_2\geq0$) diagnoses earlier a lack of representability
compared to the $D$-condition (i.e.\ $\hat\rho_2\geq0$). For both
operators, the falling of an eigenvalue below zero is accompanied by an avoided
crossing which involves the next-larger eigenvalue (this is hardly visible 
in the case of the $\hat K_2$ eigenvalue where the avoided crossing happens at 
about $t\sim71.5/J$). In fact, we have observed that level-repulsion ``pushes'' eigenvalues below zero 
in various other situations (see also Section
\ref{sec:breath}).

Now turning to the minimal invasive correction algorithms based on the $2$-norm minimization
of the correction operator $\hat{\mathcal{C}}_2$,
we set the threshold $\epsilon$ below 
which an eigenvalue is regarded as negative to $-10^{-10}$. Let us first inspect the dimensionality
of the optimization problem underlying both our purification algorithm
of the $2$-RDM and the correction algorithm of its EOM (see Appendix H 
of \cite{BBGKY_1} for the details).
The bosonic hermitian correction operator $\hat{\mathcal{C}}_2$ can be parametrized by
$m^2(m+1)^2/4=9$ real-valued parameters. Requiring $\hat{\mathcal{C}}_2$ to be contraction-free
and energy-conserving imposes $m^2+1=5$ constraints such that the system of linear
equations corresponding to the constraints is underdetermined as long
as the numbers of negative $\hat\rho_2$ eigenvalues $d$ and negative $\hat K_2$ eigenvalues
$d'$ obey $d+d'<4$.

Figure  \ref{fig:BH_correction_alg} (b.1) and (b.2) depict the results if 
the iterative minimal-invasive purification algorithm is applied after each $\Delta t=0.1/J$. 
Clearly, we see that this correction
algorithm induces strong noise in the $\hat K_2$ eigenvalues when the smallest
eigenvalue $\xi_i$ has reached significant negative values in the
uncorrected BBGKY calculation [see subfigure (a.2)]. Actually, after $t=86.5/J$, the 
iterative purification algorithm fails to converge after the maximal number of $500$ steps. 
Thus, this iterative scheme 
fails to prevent that smallest eigenvalue is pushed
to negative values due to level repulsion.

In a certain sense, we may view the iterative purification algorithm of the $2$-RDM 
as being based on a fixed stepsize as well as perturbative. In each iteration step namely, 
we update $\hat\rho_2(t)
\rightarrow\hat\rho_2(t)+\hat{\mathcal{C}}_2$ with $\hat{\mathcal{C}}_2$ 
shifting negative eigenvalues to zero in first-order perturbation theory. In the correction
algorithm for the $2$-RDM EOM, we effectively allow for variable update stepsizes by
coupling the correction scheme to the integration of the EOM, i.e.\ to the
employed integrator ZVODE \cite{zvode} featuring adaptive stepsizes. Moreover, by imposing constraints 
on the time-derivative of negative eigenvalues, we realize a non-perturbative correction
scheme.

This can nicely be inferred from the insets of Figure \ref{fig:BH_correction_alg} (c.1) and (c.2)
showing a close-up of slightly negative eigenvalues. These are exponentially damped to zero, namely
as e.g.\ $\xi_i(t+\tau)=\xi_i(t)\exp[-\eta\tau]$ for $t$ and $\tau$ such that $\xi_i(t+\tau)<\epsilon$,
with the chosen damping constant $\eta=10J$. As a consequence, the truncated BBGKY EOM 
becomes stabilized and we have observed that the $D$- and $K$-representability condition
are fulfilled to a good approximation for at least $t\leq1000/J$ (times
later than $t=150/J$ not shown in Figure \ref{fig:BH_correction_alg}).
When enforcing negative eigenvalues to be damped to zero, one might 
fear that eigenvalues accumulate in the range $[\epsilon,0]$. This, however, is not the 
case as shown in the insets of Figure \ref{fig:BH_correction_alg} (c.1) and (c.2)
because no constraint on the time-derivative of an eigenvalue is enforced if its value exceeds the threshold $\epsilon$
such that the (corrected) EOM may lift this eigenvalue above zero. We finally
remark that the number of integrator steps per $\Delta t$
significantly increases
in the vicinity of avoided crossings of 
$\hat\rho_2$ or $\hat K_2$ eigenvalues close to zero 
[see Figure
\ref{fig:BH_correction_alg} (d)]. This finding confirms the non-perturbative, adaptive 
nature of the EOM correction algorithm and at the same time highlights 
the significance of controlling such avoided crossings for a successful stabilization of the
truncated BBGKY EOM.

Without showing additional graphical illustrations, let us now briefly comment on the behavior of the correction algorithms for truncation
orders $\bar o>2$, using the Mazziotti ansatz \cite{mazziotti_purification_2002} for the correction operators $\hat{\mathcal{C}}_o$
on orders $o>2$ (see Section V B 
of \cite{BBGKY_1}).
Focusing first on the RDM purification, we have observed that $\hat\rho_{\bar o}$
can be kept positive semi-definite up to a few tens $1/J$ longer (compared to the uncorrected case) before
this iterative correction algorithm fails to converge after $500$ steps. Due 
to the losing of positive semi-definiteness in decreasing sequence
with respect to the RDM order (see Figure \ref{fig:BH_neg_t}), 
we found for $\bar o\geq 4 $ that also  $\hat\rho_{\bar o-1}\geq0$ 
is valid for somewhat longer times compared to the
uncorrected case. Unfortunately, however, this correction scheme fails to converge so
early that it does not improve the timescale, on which 
the most important RDMs for making predictions for ultracold quantum gas
experiments, namely $\hat\rho_1$ and $\hat\rho_2$, obey the considered representability 
conditions. 

Extending the EOM correction scheme to higher truncation orders $\bar o>2$
by means of the Mazziotti ansatz for the higher-order correction operators unfortunately
proved to be quite unsuccessful. This failure manifests itself in an enormous increase
of integrator steps per $\Delta t$, i.e.\ the EOM becoming stiff, in combination
with the quadratic optimization problem for determining $\hat{\mathcal{C}}_2$ having
no solution, i.e.\ constraints contradicting one another. Unfortunately, we
cannot tell whether the latter is a fundamental problem or whether it is 
only induced by the EOM to become stiff 
due to an inappropriate ansatz of $\hat{\mathcal{C}}_o$ for $o>2$, potentially
leading to integration errors.

To sum up, while we can successively stabilize the BBGKY EOM  truncated at
order $\bar o=2$ by enforcing the $D$- and $K$-representability condition
via a minimal invasive correction of the $2$-RDM EOM, the issue of 
higher-order correlations becoming dominant after a certain time remains unsolved
in this example. Since this tunneling scenario might well be
unsuitable for a closure approximation based on neglecting certain few-particle correlations,
we now turn to an example, where a BEC becomes only slightly depleted in the course of the
quantum dynamics.
\section{Interaction-quench induced breathing dynamics of harmonically trapped bosons}\label{sec:breath}

In this application, we are concerned with collective excitations of
$N$ ultracold bosons confined to a quasi one-dimensional harmonic trap. In harmonic
oscillator units (HO units), the corresponding Hamiltonian reads
\begin{align}
 \hat H = \sum_{i=1}^N\frac{\hat p_i^2+\hat x_i^2}{2}+g\sum_{1\leq i<j\leq N}\delta(\hat x_i-\hat x_j)
\end{align}
where we model the short-range van-der-Waals interaction by the contact potential \cite{Pethick_Smith2008}
of strength $g$. Initially, we assume all atoms to reside in the ground state of 
the single-particle Hamiltonian, i.e.\ a Gaussian orbital, which
is the exact many-body ground state in the absence of interactions. Then, the interaction strength
is instantaneously quenched to $g=0.2$ such that the ideal BEC becomes slightly 
depleted and its density performs breathing oscillations, i.e.\ expands and contracts periodically.
This so-called breathing mode has been investigated theoretically as well as experimentally in different settings (see e.g.\ \cite{Stringari2002,Moritz2003,bauch_2009,Bonitz2014,Schmitz2013,Tschischik2013,Bouchoule2014} for single-component
systems
and e.g.\ \cite{pyzh_spectral_2017} for mixtures), and measuring its frequency proves to be useful
for characterizing the interaction regime \cite{Haller2009}.

\begin{figure}[t]
 \includegraphics[width=0.495\textwidth]{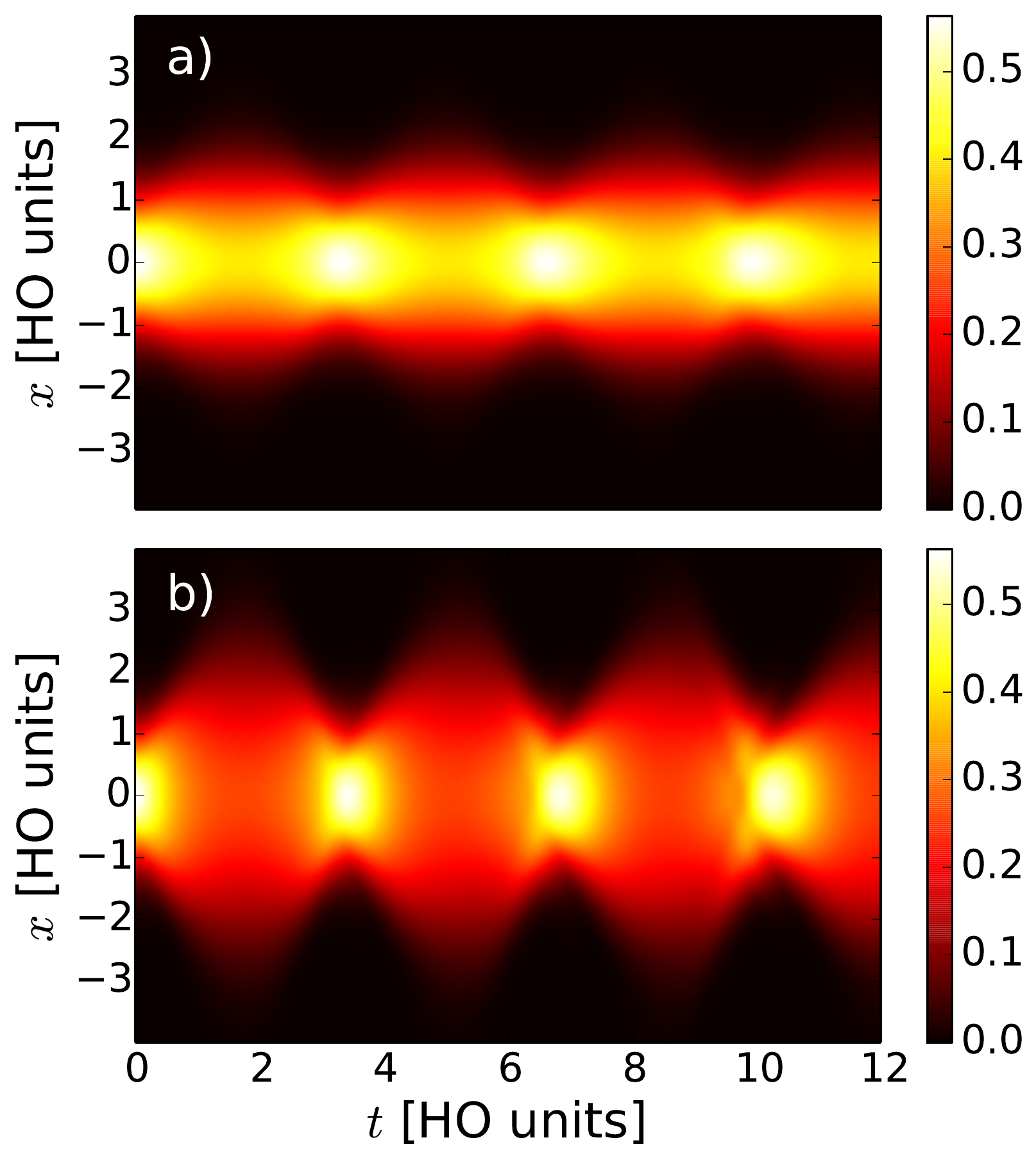}
 \caption{(color online) Time-evolution of the reduced one-body density $\rho_1(x;t)=\langle x|\hat \rho_1(t)| x\rangle$
 for $N=10$ [subfigure (a)] and $N=30$ [subfigure (b)] bosons quenched from the non-interacting ground-state
 to a contact-interaction strength of $g=0.2$. These results are obtained by MCTDHB simulations with $m=4$ dynamically
 optimized SPFs.}
  \label{fig:breath_dens_ex}
\end{figure}

Before we discuss the results of the truncated BBGKY approach, let us first
inspect the results of MCTDHB simulations with $m=4$ dynamically optimized SPFs,
which shall serve as the reference in the following. For representing the SPFs,
a harmonic discrete variable representation \cite{dvr_and_their_utilization_Light_Carrington_2000,MCTDH_BJMW2000} with $n=256$ ($n=320$) grid points
is employed for case of $N=10$ ($N=30$) particles.
In Figure \ref{fig:breath_dens_ex},
we depict the time evolution of the reduced one-body density, i.e.\ the diagonal 
of the $1$-RDM in position representation  $\rho_1(x;t)=\langle x|\hat \rho_1(t)| x\rangle$,
for $N=10$ and $N=30$ bosons. In both cases, we clearly see that the atomic density 
periodically expands and contracts. Since the interaction quench leads 
to a more than three times larger interaction energy per particle of the ensemble of $N=30$ atoms
compared to $N=10$ (at $t=0$), the density of the former expands much further
into the outer parts of the trap. In contrast to this, the density of the $N=10$ atom
ensemble seems to stay Gaussian (with a time-dependent width) to a good approximation,
indicating that we operate in the linear-response regime here. We note 
that the quench leads to only a slight quantum depletion of at most $3\%$ in both cases (see below).

In the following, we first show that the truncated BBGKY approach leads to stable
results in the $N=10$ case, whose accuracy can be systematically improved by
increasing $\bar o$. Thereafter, we turn to the $N=30$ case where
we again encounter instabilities of the EOM and thus apply correction algorithms.
We stress that for both
cases we  operate with $m=4$ dynamically optimized SPFs,
solving the truncated BBGKY EOM coupled to the MCTDHB EOM for the SPFs, 
which is in contrast to the Bose-Hubbard tunneling scenario of Section
\ref{sec:BH_dimer}.

\subsection{Breathing dynamics of $N=10$ bosons}

\begin{figure}[t]
   \includegraphics[width=0.495\textwidth]{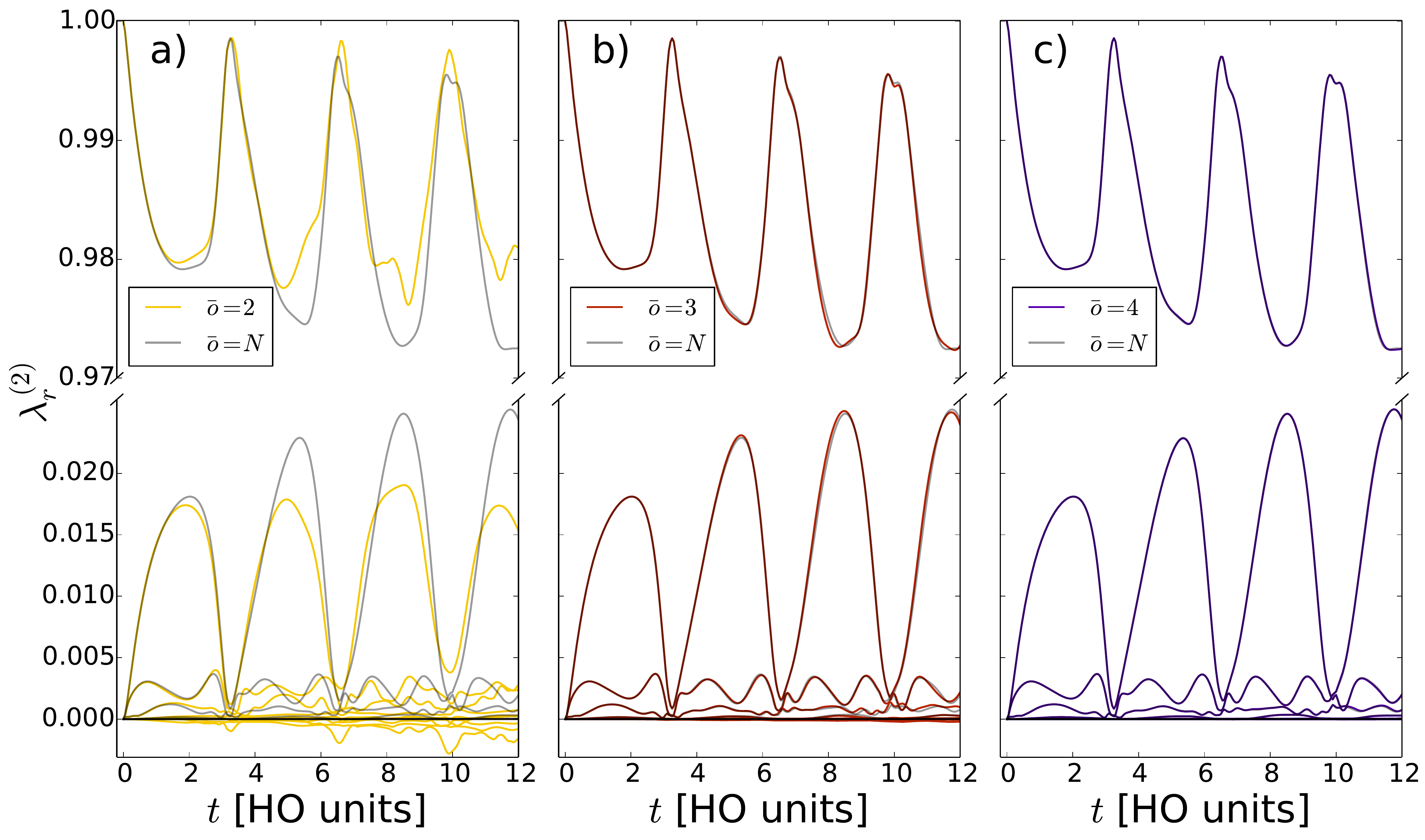}
   \caption{(color online) Natural populations of the $2$-RDM for 
   the truncation orders $\bar o=2$ [a)], $\bar o=3$ [b)] and $\bar o=4$ [c)]
   in comparison to the \mbox{MCTDHB} results. Parameters: $N=10$ atoms, post-quench interaction
   strength $g=0.2$, $m=4$ dynamically
   optimized SPFs.}
   \label{fig:breathing_NP2_N10}
\end{figure}

In Figure \ref{fig:breathing_NP2_N10}, we show the time-evolution of the 
$2$-RDM NPs for various truncation orders. Focusing first on the MCTDHB results,
we see that correlations (in the sense of deviations from a Gross-Pitaevskii mean-field
state where on all orders $o$ 
there is only one finite NP $\lambda_1^{(o)}=1$ and all other NPs vanish) repeatedly emerge and decay. The deviations from the NP distribution
of a Gross-Pitaevskii mean-field state is approximately most pronounced when the 
density is most spread-out and become almost negligible when the density has approximately recovered
its initial width [see Figure \ref{fig:breath_dens_ex} a)].

While the truncated BBGKY results for $\bar o=2$ feature significant deviations
from the MCTDHB results, the results drastically improve when going to $\bar o=3$
and become practically indistinguishable from the MCTDHB results already
at the truncation order $\bar o=4$. Actually, convergence of the $1$-RDM NPs 
$\lambda^{(1)}_i$ is reached even at $\bar o=3$ (not shown). Coming back to $\bar o=2$,
we point out that the $2$-RDM quickly becomes indefinite where small negative eigenvalues
are in particular pushed further to larger negative values when the density
contracts to its initial width and small but positive NPs approach zero. As in 
the case of the above tunneling scenario, we interpret this finding as ``induced'' by
level repulsion. Upon increasing the truncation order, we see that the $2$-RDM 
stays positive semi-definite on the considered time-interval, which is a nice 
example for how increasing the accuracy of the closure approximation
also stabilizes the truncated BBGKY EOM.

\begin{figure*}[t]
 \includegraphics[width=0.495\textwidth]{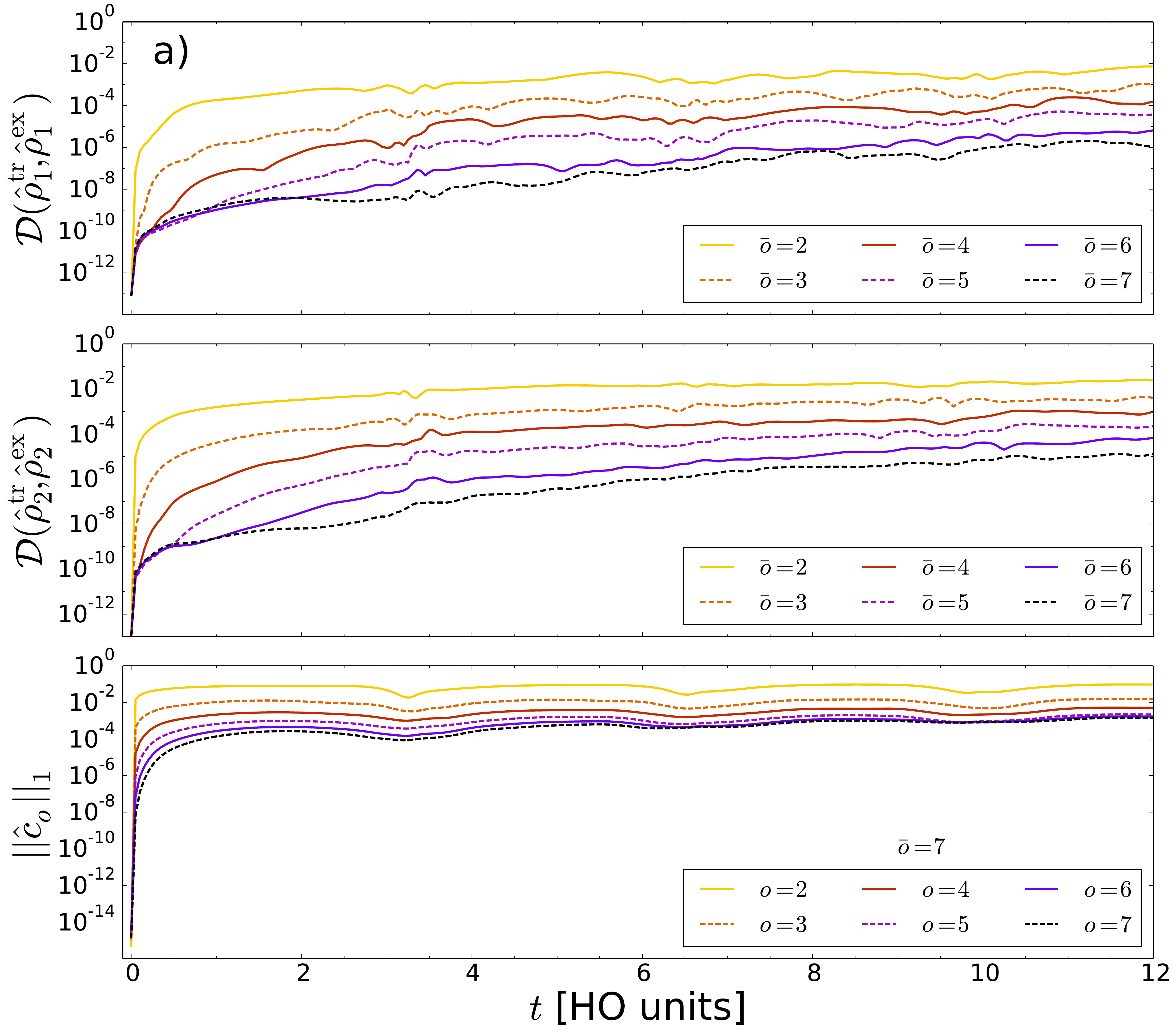}
  \includegraphics[width=0.495\textwidth]{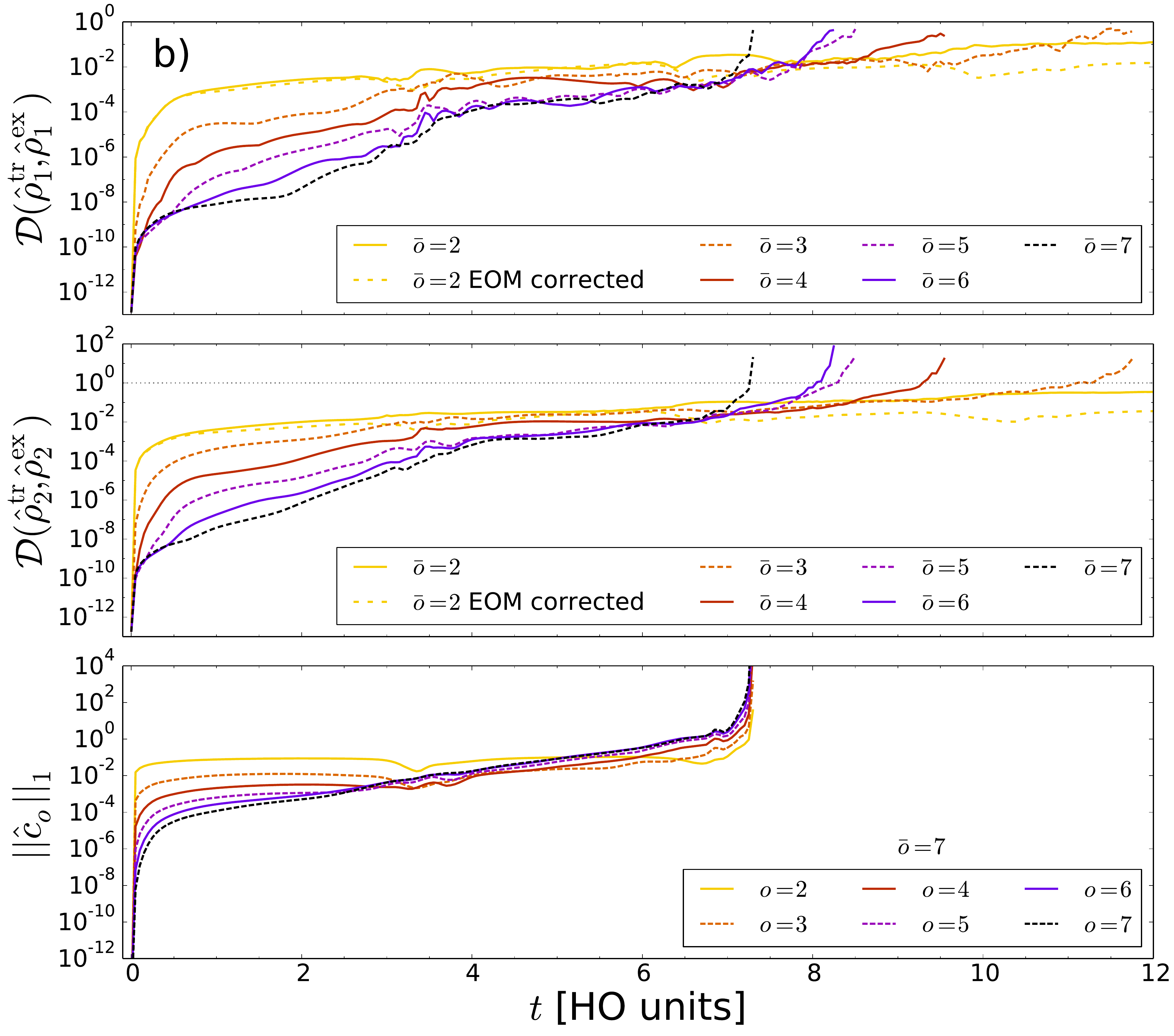}
 \caption{(color online) 
 First and second row:
 Time evolution of the trace distance $\mathcal{D}(\hat\rho_o^{\rm tr},\hat\rho_o^{\rm ex})$
 between the MCTDHB and the truncated BBGKY prediction for the $o$-RDM ($o=1,2$) and various truncation
 orders $\bar o$. The dotted horizontal line at unity ordinate value indicate the upper bound for the trace distance
 between two density operators. Third row: Time-evolution of the cluster's trace-class norm $||\hat c_o||_1$
 for $o=1,...,7$ obtained from the data of the $\bar o=7$ simulations. Left column: $N=10$ atoms. Right columns: $N=30$.
 Otherwise, same parameters as in Figure \ref{fig:breathing_NP2_N10}.}
 \label{fig:breathing_rdm_comp}
\end{figure*}

For a systematic comparison, we next compare the trace-class distance
$\mathcal{D}(\hat\rho_o^{\rm tr},\hat\rho_o^{\rm ex})$
between the truncated BBGKY result for the $o$-RDM, $\hat\rho_o^{\rm tr}$, and 
the corresponding MCTDHB result, $\hat\rho_o^{\rm ex}$, in Figure 
\ref{fig:breathing_rdm_comp} a). We remark that although the SPFs of the truncated BBGKY approach 
obey the same EOM (5) 
of \cite{BBGKY_1} as the
dynamically optimized SPFs of the MCTDHB method, we cannot expect these two
sets of SPFs to coincide because the $1$- and $2$-RDM entering
the SPF EOM differ in general, which has to be taken into account
when calculating $\mathcal{D}(\hat\rho_o^{\rm tr},\hat\rho_o^{\rm ex})$.
In stark contrast to the tunneling scenario, we see that the accuracy of the
truncated BBGKY results for the $1$- and $2$-RDM 
systematically improves upon increasing $\bar o$ for all considered times.

Finally, we quantify the strength of few-particle correlations
in terms of $||\hat c_o||_1$, as extracted from the $\bar o=7$ calculation [see
Figure \ref{fig:breathing_rdm_comp} a)]. Here, we see that the correlations 
stay bounded on the considered time interval and are ordered in a
clear hierarchy, i.e.\ $||\hat c_{o+1}||_1(t)<||\hat c_{o}||_1(t)$.
Apparently, these are ideal working conditions for the truncated BBGKY approach.

\subsection{Breathing dynamics of $N=30$ bosons}
Next, let us increase the quench-induced excitation energy per particle by more than
a factor of three when going to $N=30$ bosons and keeping the post-quench 
interaction strength $g=0.2$ the same. Similarly to the tunneling scenario,
we first inspect the natural populations, then compare lowest order RDMs and 
finally evaluate the performance of the correction algorithms under discussion.

\subsubsection{Natural populations}

\begin{figure*}[t]
 \includegraphics[width=0.9\textwidth]{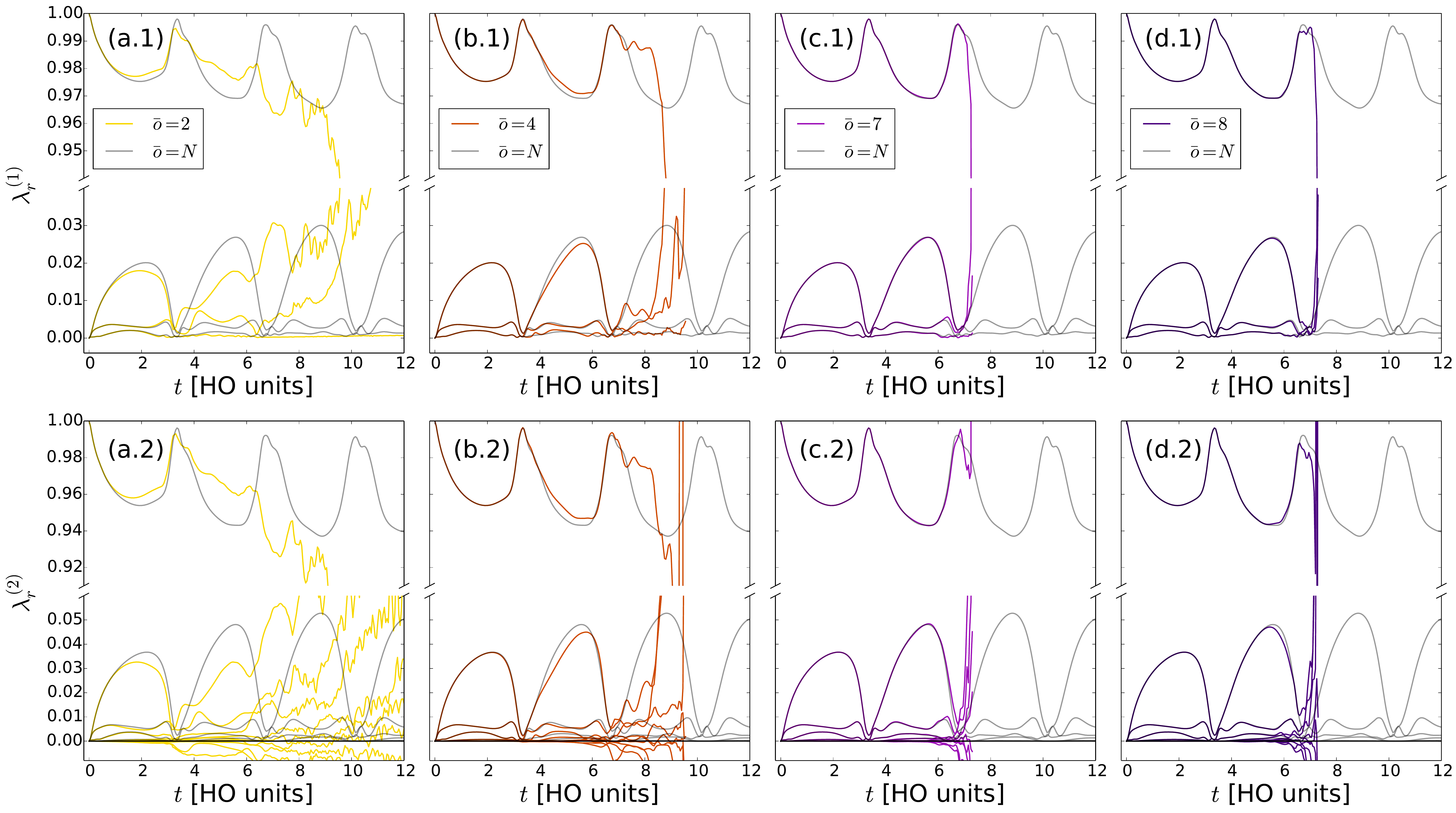}
 \caption{(color online) Top (bottom) row: Time-evolution of the $1$-RDM ($2$-RDM) NPs 
 for various truncation orders $\bar o$ in comparison to the MCTDHB results. 
 We note that the ordinates are broken into two parts for covering the whole 
 range of relevant values. In some cases, this leads to discontinuous curves, see
 e.g.\ the $\bar o=2$ curve in (a.2).
 Number
 of bosons: $N=30$.
  Otherwise, same parameters as in Figure \ref{fig:breathing_NP2_N10}.}
  \label{fig:breathing_NPs_N30}
\end{figure*}

In Figure \ref{fig:breathing_NPs_N30}, we show the NPs of the $1$- and $2$-RDM for various
truncation orders $\bar o$ in comparison
to the MCTDHB results. Similarly to the $N=10$ case, we see how the NP distributions as obtained
from MCTDHB
oscillates between the characteristics of the Gross-Pitaevskii mean-field state
and a (slightly) correlated one, which is approximately synchronized to the strongest
contraction and expansion of the density, respectively [see Figure 
\ref{fig:breath_dens_ex} b)]. In contrast to the former case, however,
we can converge the NPs to the MCTDHB results upon increasing the
truncation order $\bar o$ only for times $t\lesssim 5$ HO units. 
For all considered truncation
orders, we witness an exponential-like instability in the $2$-RDM NPs resulting in large negative
eigenvalues while the $1$-RDM stays positive semi-definite for the 
considered time-span. Fixing $\bar o$, we have observed also for this scenario that 
the lack of positive semi-definiteness of the $o$-RDMs  happens in 
decreasing sequence with respect to the order $o$ (not shown). 
Moreover,
these instabilities in the $2$-RDM NPs seem to be triggered by small positive NPs approaching zero
from above, namely when the density approximately shrinks to its initial width,
see e.g.\ Figure \ref{fig:breathing_NPs_N30} (c.2).
Finally, we have observed for the case $\bar o=2$ that increasing the number of SPFs from $m=4$ to $m=8$
slightly enhances the time-scale on which the instability of the $2$-RDM NPs 
takes place (not shown). This finding is reasonable since the
projector $(\mathds{1}-\hat{\mathbb{P}})$, occurring in the SPF EOM (5) 
of \cite{BBGKY_1},
projects onto a smaller subspace when increasing $m$ such that the impact
of the non-linearity in the SPF EOM is effectively reduced.

\subsubsection{Reduced density operators}
Comparing the BBGKY prediction for the complete $1$- and $2$-RDM with the corresponding MCTDHB results in terms of the trace-class distance in Figure \ref{fig:breathing_rdm_comp} b),
we see that deviations emerge much faster as compared to the $N=10$ case.
 At longer times, we also
observe that the accuracy of the BBGKY results does not monotonously increase anymore
with increasing $\bar o$. Moreover, the above mentioned instabilities 
also partly manifest themselves in $\mathcal{D}(\hat\rho_2^{\rm tr},\hat\rho_2^{\rm ex})$
attaining unphysical values above unity. Finally, we also
depict $||\hat c_o||_1$ as a measure for correlations in Figure 
\ref{fig:breathing_rdm_comp} b). While the correlations
are hierarchically ordered in decreasing sequence with respect to the order $o$ up
to $t\sim 2.7$ HO units, this ordering becomes  reversed later on.
This finding, however, is not conclusive, i.e.\ might be unphysical and 
related to the observed instability, since the values of $||\hat c_o||_1$ have been extracted 
from the BBGKY data with $\bar o=7$ (in contrast to the tunneling 
scenario where the numerically exact $\hat c_o$ have been used).

At this point, we shall remark that we expect a much better agreement for the
$N=30$ case when quenching to much lower interaction
strengths $g\ll0.2$ and thereby reducing the overall excitation energy.

\subsubsection{Performance of the correction algorithms}
\begin{figure*}[t]
 \includegraphics[width=0.9\textwidth]{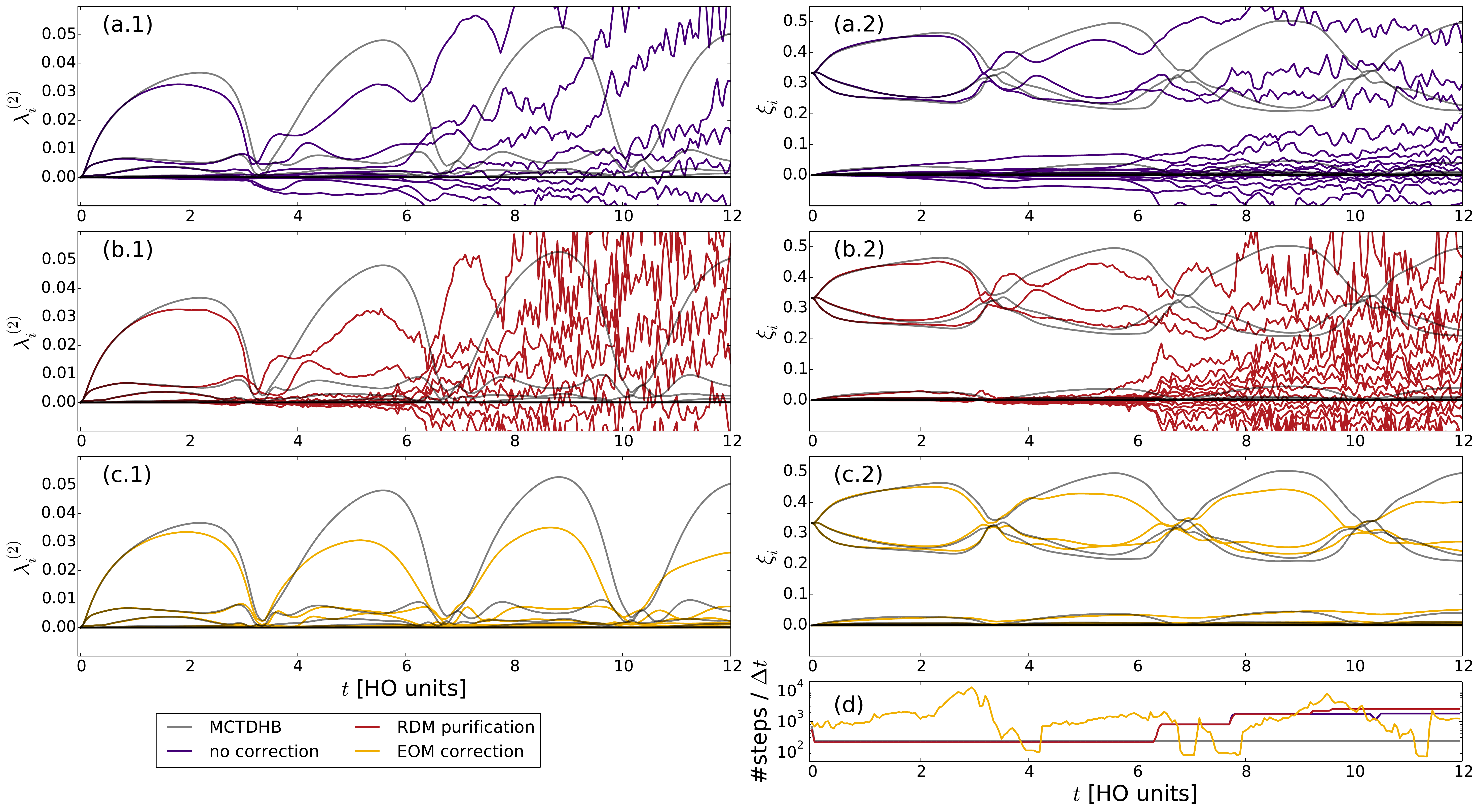}
 \caption{(color online)  
 Comparison of the correction
 strategies outlined in Section V B 
 of \cite{BBGKY_1} (for the BBGKY hierarchy
 truncated at $\bar o=2$), i.e.\ same as Figure \ref{fig:BH_correction_alg}
 but for the interaction-quench scenario with $N=30$ bosons. Parameters:
 threshold $\epsilon=-10^{-10}$, damping constant $\eta=10$ HO units, write-out
 time-step $\Delta t=0.05$ HO units, maximal number of iterations: $500$. 
  Otherwise, same parameters as in Figure \ref{fig:breathing_NP2_N10}.}
  \label{fig:breathing_correction_alg}
\end{figure*}

Here, we again focus mainly on the performance of the correction algorithms
applied to the $\bar o=2$ BBGKY approach and comment later on larger truncation 
orders. In Figure \ref{fig:breathing_correction_alg}, we depict the spectrum\footnote{
We note that the $\hat K_2$ spectra at $t=0$ in the tunneling and the interaction-quench 
scenario differ although the system is initially a fully condensed BEC in both cases
(see Figures \ref{fig:BH_correction_alg} and \ref{fig:breathing_correction_alg}).
This is due to the fact that the $\hat K_2$ spectrum is sensitive to the total number
of SPFs $m$ even if not all of them are occupied.} of
$\hat K_2$ as well as a close-up to the $2$-RDM spectrum in the vicinity of zero
for the uncorrected BBGKY approach, the minimal invasive RDM purification algorithm
and the minimal invasive EOM correction algorithm. In the minimization problem
underlying both correction algorithms, we have to find the optimal
$\hat{\mathcal{C}}_2$ which depends on $m^2(m+1)^2/4=100$ real-valued parameters.
Being contraction-free and energy conserving leads to $m^2+1=17$ constraints. 
Moreover, $\hat{\mathcal{C}}_2$ has to obey the parity symmetry of our problem
imposing $m^4/8+m^3/4=48$ further constraints (see Appendix H 
of \cite{BBGKY_1}). Thereby,
our system of linear constraints remains underdetermined as long 
as the number $d$ of negative $\hat\rho_2$ eigenvalues and number $d'$ of negative
$\hat K_2$ eigenvalues obey $d+d'<35$.

In Figure \ref{fig:breathing_correction_alg}, we see that the minimal-invasive
RDM purification algorithm clearly suppresses significant negative eigenvalues until
$t\sim2.5$ HO units. Thereafter, noticeably negative eigenvalue emerge but stay
bounded from below until $t\sim6$ HO units when a drastic
instability kicks in. Thus, this iterative algorithm soon fails to converge
after the maximal number of $500$ iteration steps. In order to understand
the deeper reason of this failure, we have analyzed the spectrum
of the updated operators $\hat\rho_2(t)+\alpha\,\hat{\mathcal{C}}_2$
and $\hat K_2(t)+\alpha\,\hat{\Delta}_2$
for $\alpha\in[0,1]$ and the first few iteration steps at such an instant in time (not shown). Thereby, we
have found that while the tangent on a negative eigenvalue (with respect to $\alpha$)
indeed crosses zero as imposed by our constraints, level repulsion 
with other (in most cases negative) eigenvalues often hinders this negative eigenvalue
to significantly move towards zero. We cannot rigorously prove that this is indeed
the only mechanism for the breakdown of this iterative purification algorithm, of course.

Yet at least, this finding gives a useful hint why our non-perturbative, adaptive approach,
 the minimal invasive correction scheme of the $2$-RDM EOM, gives very
 stable results [see Figure \ref{fig:breathing_correction_alg} (c.1) and (c.2)].
 Actually, we observe that the $D$- and $K$-conditions are fulfilled to
 a good approximation much longer, namely for at least $t\leq 36$ HO unit (not shown).
 From Figure \ref{fig:breathing_correction_alg} (d), we furthermore infer that the integrator
 variably adapts its step-size, but in contrast to the Bose-Hubbard tunneling scenario
 no systematic enhancement of integrator steps is observed when $\hat\rho_2$ or
 $\hat K_2$ eigenvalues avoid each other in the vicinity of zero. Apparently, the
 though stabilized result features noticeable deviations from the MCTDHB
 results for the respective eigenvalues.
 Yet, we find that the overall accuracy of the $\bar o=2$ results for the $1$-
 and $2$-RDM as measured by the trace-class distance is systematically
 improved for most times by correcting the $2$-RDM EOM, 
 as one can infer from Figure \ref{fig:breathing_rdm_comp} b).

\begin{figure}[t]
 \includegraphics[width=0.495\textwidth]{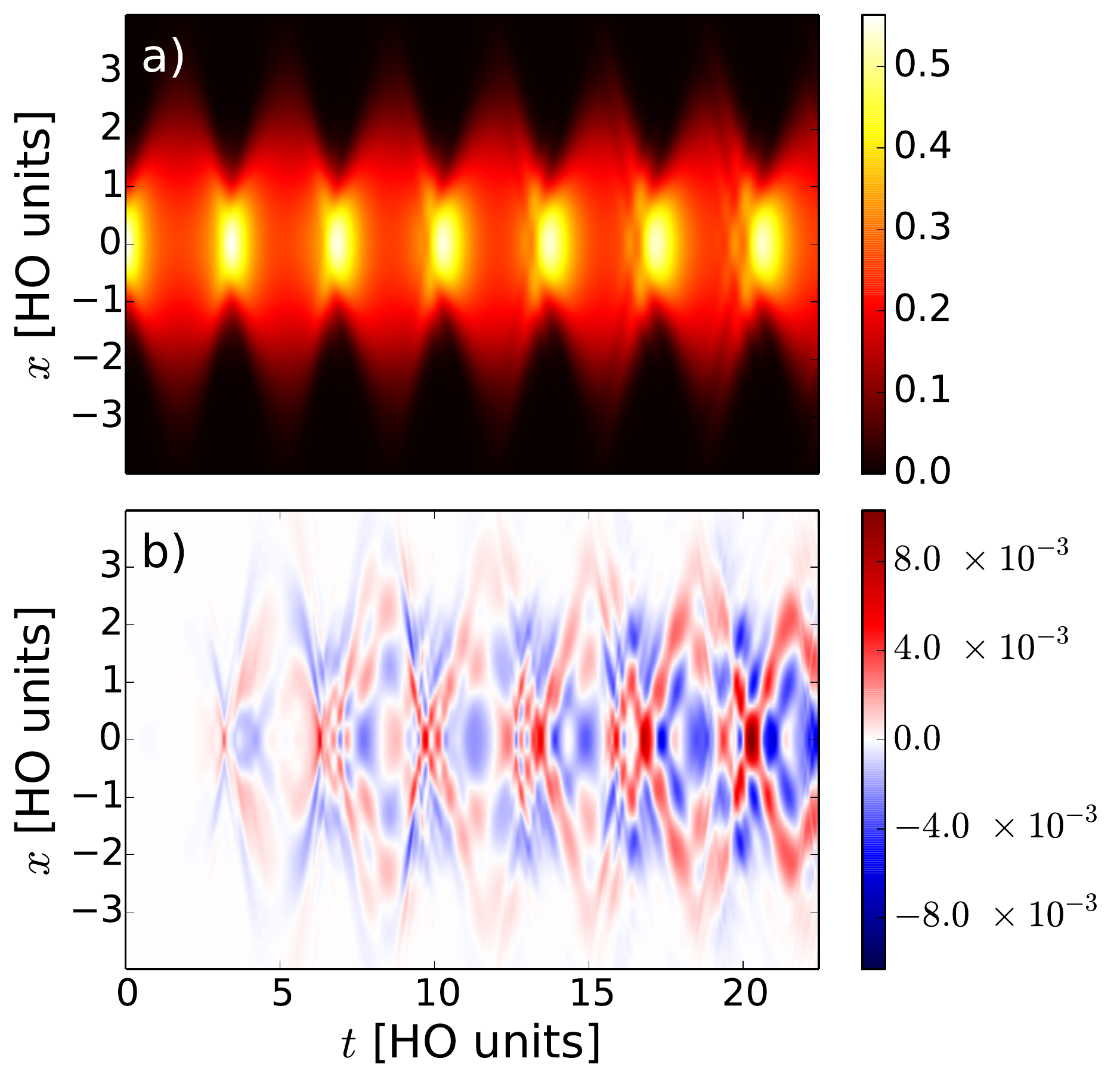}
 \caption{(color online) Subfigure a): Time-evolution of the reduced one-body density $\rho_1^{\rm ex}(x;t)$
 as obtained from \mbox{MCTDHB} on a longer time-scale. 
 Subfigure b): Absolute deviation $\rho_1^{\rm tr}(x;t)-\rho_1^{\rm ex}(x;t)$
 of the reduced one-body density $\rho_1^{\rm tr}(x;t)$ 
 as obtained from the BBGKY approach truncated at $\bar o=2$ and 
 stabilized by the minimal-invasive EOM correction algorithm with
  $\epsilon=-10^{-10}$ and $\eta=10$ HO units. Number of bosons: $N=30$. 
  Otherwise, same parameters as in Figure \ref{fig:breathing_NP2_N10}.}
  \label{fig:breathing_dens_comp_N30}
\end{figure}

 In order to judge the accuracy of the EOM-corrected $\bar o=2$ simulation
more descriptively, we depict the deviations of its prediction for the 
reduced one-body density from the MCTDHB results in Figure \ref{fig:breathing_dens_comp_N30}.
Note that this plot covers a longer time-span compared to the previous ones.
As expected, we find that the deviations increase in time. Compared to the absolute values of 
the density, these deviations are, however, small and, most importantly, somewhat
smaller than the deviations of corresponding Gross-Pitaevskii mean-field simulation from
the $m=4$ MCTDHB results (not shown). Finally, let us connect
the errors in the one-body density to the errors measured by the 
trace-class distance $\mathcal{D}(\hat\rho_1^{\rm tr},\hat\rho_1^{\rm ex})$
as depicted in Figure \ref{fig:breathing_rdm_comp} b). For this purpose, we note that
the density at position $x$ can be expressed as the expectation value
of the one-body observable $\hat A_1=|x\rangle\!\langle x|$ featuring $||\hat A_1||_1=1$.
Thereby, the inequality of Appendix \ref{app:exp_value_bound}
gives $|\rho_1^{\rm tr}(x;t)-\rho_1^{\rm ex}(x;t)|\leq 2
\mathcal{D}(\hat\rho_1^{\rm tr}(t),\hat\rho_1^{\rm ex}(t))$,
which is consistent with the results depicted in Figure \ref{fig:breathing_dens_comp_N30}.

Going to higher truncation orders by making the Mazziotti ansatz for the 
corresponding higher-order correction operators $\hat{\mathcal{C}}_o$ unfortunately does
not improve the BBGKY results, as already observed in the tunneling scenario.
While the iterative RDM purification 
scheme fails to prevent the instabilities, we observe the same obstacle 
for the EOM correction algorithm as previously encountered, namely the optimization problem
at order $o=2$ lacking a solution (results not shown). Yet due to the very promising
results of the EOM correction algorithm when truncating the BBGKY hierarchy at order
$\bar o=2$, we believe that extending the EOM correction algorithm to higher orders
without employing the Mazziotti ansatz for the correction operator
is a highly promising direction to go.

\section{Conclusions}\label{sec:concl}
While we have provided the theoretical framework of the truncated BBGKY hierarchy for
simulating the quantum dynamics of few-particle reduced density operators of 
an ultracold bosonic many-body system in the immediately preceding work \cite{BBGKY_1},
we have provided here extensive applications.  The central aim of this work is to systematically study 
the impact of the  BBGKY hierarchy truncation order $\bar o$ on the numerically obtained results
where the highly efficient formulation of the theory in \cite{BBGKY_1} allows us for going
to truncation orders as high as $\bar o=12$. For this purpose, we have
considered both a tunneling scenario where a Bose-Einstein condensate fragments into
a two-fold fragmented, highly entangled state and thus strong correlations emerge, and 
an interaction-quench scenario where a Bose-Einstein condensate becomes weakly depleted
when performing collective oscillations. While the former scenario is based on the Bose-Hubbard 
dimer model where we may use
time-independent Wannier states for representing the reduced density operators (RDMs), we 
solve in the second scenario the truncated BBGKY hierarchy as derived from the variational Multi-Configuration Time-Dependent Hartree method
for Bosons (MCTDHB) \cite{BBGKY_1,MCTDHB_PRA08} and thereby employ dynamically
optimized single-particle functions for efficiently representing the RDMs.
The truncated BBGKY results are comprehensively evaluated by inspecting a variety
of observables, covering descriptive ones such as the atomic density but also the eigenvalues
of the low-order RDMs and their trace-class distances
to the corresponding numerically exact and \mbox{MCTDHB} results in the first and second scenario,
respectively.

In all applications, we have found that the short-time dynamics can be highly accurately
described by the truncated BBGKY approach, where the accuracy of the results systematically
improves with increasing truncation order $\bar o$. At longer times, the BBGKY gives also excellent 
results with controllable accuracy for the interaction-quench scenario and not too high
excitation energies. For higher excitations energies as well as in the tunneling scenario,
however, we obtain only faithful results up to a certain time even if we 
truncate the hierarchy at the highest non-trivial order $\bar o=N-1$ with $N$ denoting
the total number of particles. After that time, the accuracy does not monotonously improve
with increasing truncation order and instabilities emerge, manifesting
themselves in an exponential-like decrease of RDM eigenvalues to negative values.
By inspecting the exact numerical results for the tunneling scenario,
we find that few-particle correlations on all orders quickly
play a significant role and eventually $N$-particle correlations dominate 
because the total system evolves into a NOON-state. This finding indicates
that the long-time physics of this scenario prevents to use a truncation approximation
which is based on neglecting $(\bar o+1)$-particle correlations.

Nevertheless, it is important to clearly separate the stability properties of the 
truncated BBGKY equations of motion (EOM) from accuracy issues because (i) it is not
desirable to have a highly accurate theory which is exponentially unstable under slight e.g.\
numerical perturbations and (ii) also a not highly accurate truncation approximation
may give useful, sufficiently accurate results for low-dimensional observables such as the 
density if the EOM are sufficiently stable. In order to disentangle stability issues
from the accuracy of the results, we have performed an in-depth analysis of the observed exponential-like
instabilities. In all examples, we witness that the instability sets in at the truncation 
order $\bar o$ and then propagates down to lower orders meaning that $o$-particle RDMs lack 
to be positive semi-definite in decreasing sequence with respect to the order $o$. The time until 
which the highest-order RDM, $\hat \rho_{\bar{o}}$, stays positive 
semi-definite only gradually increases with the truncation order $\bar o$.
Upon increasing $\bar o$, the most important RDMs for calculating relevant observables for ultracold systems, $\hat\rho_1$ and $\hat\rho_2$,
are found to stay a bit longer positive semi-definite in some examples, in
others, however, the opposite holds. That the instabilities set in more rapidly with increasing
truncation order $\bar o$ might by explained by the fact that the cluster-expansion
based truncation approximation (see Section I B 3 
of \cite{BBGKY_1}) constitutes a polynomial of degree $(\bar o+1)$ in $\hat\rho_1$
and of degree $\lfloor(\bar o+1) /o\rfloor$ in the cluster $\hat c_o$. Thereby,
the non-linearity of the EOM is effectively enhanced for increasing $\bar o$.
Moreover, the instabilities often take
place when the lowest RDM eigenvalues undergo an avoided crossing in the vicinity of zero.

Ultimately, these instabilities emerge due to the applied truncation
approximation, which conserves the RDM compatibility and energy as well as respects
symmetries such as the parity invariance if existent, but does not ensure
that important necessary representability conditions such as the positive
semi-definiteness of $\hat\rho_o$ and of the one-particle-one-hole RDM 
are fulfilled \cite{BBGKY_1}. In order to enforce such representability conditions
and stabilize the EOM for a given truncation order $\bar o$ (independently of the
accuracy of the results), we have evaluated the two novel correction algorithms 
developed in \cite{BBGKY_1}. Both strategies aim at ensuring the positive semi-definiteness
of $\hat\rho_2$ and the (modified) one-particle-one-whole RDM as defined in \cite{garrod_1964}
in a minimal-invasive and energy-conserving way by solving an optimization problem.
For higher orders, the Mazziotti ansatz \cite{mazziotti_purification_2002}  
for the correction operator is chosen. The first strategy corrects the solution 
of the truncated BBGKY EOM after each small time-step $\Delta t$ by 
iteratively adding a correction operator which raises negative eigenvalues to zero
in first order perturbation theory. Unfortunately, this iterative algorithm does not 
lead to a significant stabilization over a reasonable time-span and soon fails to
converge most likely due to level-repulsion of the lowest eigenvalues which cannot 
be overcome by a perturbative approach. Yet the second approach, which corrects the truncated BBGKY EOM themselves in a 
minimal-invasive, energy-conserving way such that negative eigenvalues are exponentially 
damped to zero, has proven to stabilize the dynamics for long times if applied to the truncation order $\bar o=2$.
This highlights that a successful stabilization of the dynamics requires a non-perturbative control of 
the low-lying spectrum of e.g.\ $\hat\rho_2$. Moreover, the long-time results of the EOM-corrected BBGKY approach for various
observables are also qualitatively correct in the case of the
interaction-quench scenario. Yet
extending the EOM-correction algorithm to
higher orders by the Mazziotti ansatz has been found to lead to stiff EOM or the optimization 
problem at order $o=2$ lacking a solution. 

Due to the success for $\bar o=2$, however,
we believe that research on how to properly and efficiently extend the EOM-correction 
algorithm to higher orders constitutes a highly promising route for the future,
which can result in a robust BBGKY approach with controllable accuracy for
broad applications. A second future line of research should address how to
improve the truncation approximation of the BBGKY hierarchy for bosonic
systems with a definite number of particles, where the existing literature
is rather sparse compared to the fermionic case. 
We note here that all results of this work rely on the chosen definition of few-particle correlations,
of course.
One novel
approach for the truncation might involve to use machine-learning techniques such as artificial
neural networks
for deducing the unknown collision integral at the truncation order in
combination with an EOM correction algorithm for stabilizing the dynamics.
In any case, the theoretical framework and the highly efficient formulation
of the theory as provided in \cite{BBGKY_1} together with the deep 
empirical insights of this work serve as an important step 
in the direction of establishing the BBGKY approach as a simulation
tool for ultracold atomic systems.

\begin{acknowledgments}
The authors thank Iva B\u{r}ezinov\'a, Joachim Burgd\"orfer, David A.\ Mazziotti,
Hans-Dieter Meyer, Angel Rubio, Johannes M.\ Schurer and Jan Stockhofe for fruitful discussions.
This work has been supported by the excellence cluster ``The Hamburg Centre for Ultrafast Imaging 
- Structure, Dynamics and Control of Matter at the Atomic Scale'' of the Deutsche Forschungsgemeinschaft.
\end{acknowledgments}

\appendix
\vspace*{0.5cm}
\section{Integration of the truncated BBGKY EOM}\label{app:integration}

In both scenarios, we employ the variable-coefficient ordinary differential equation
solver ZVODE \cite{zvode} for integrating the EOM (5),(6) 
 of \cite{BBGKY_1}. 
The conservation of hermiticity of the RDMs
is numerically ensured by only propagating the lower triangle of the 
matrix-valued EOM (6) 
of \cite{BBGKY_1}, which at the same time reduces
the number of variables to be integrated. Since the applied truncation approximation
conserves the compatibility of the RDMs, we propagate only the BBGKY EOM (6) 
of \cite{BBGKY_1} at the truncation order $\bar o$ and obtain the
RDMs of lower order by partial tracing.
\section{Trace distance and expectation value of observables}\label{app:exp_value_bound}
Here, we briefly review why the trace distance $\mathcal{D}(\hat\rho_o^{\rm tr},\hat\rho_o^{\rm ex})$
constitute a bound for the difference in the prediction for the expectation value of any given
trace-class $o$-body observable $\hat A_o$. Introducing the
spectral decompositions $\hat A_o=\sum_r\alpha_r|a_r\rangle\!\langle a_r|$ and 
$\hat\rho_o^{\rm tr}-\hat\rho_o^{\rm ex}=\sum_r\beta_r|b_r\rangle\!\langle b_r|$, we
may estimate
\begin{align}
 \label{eq:proof_exp_values_and_tr_dist}
 \big| \tr\big(\hat A_o\hat\rho_o^{\rm tr}\big)- &\tr\big(\hat A_o\hat\rho_o^{\rm ex}\big)\big|
 \leq\sum_r|\beta_r|\,\big| \langle b_r|\hat A_o|b_r\rangle\big|\\\nonumber
 &\leq\sum_{r,s}|\alpha_s||\beta_r|\,\big|\langle b_r|a_s\rangle\big|^2
 \leq\sum_{r,s}|\alpha_s||\beta_r|\\\nonumber
 &=2||\hat A_o||_1\,\mathcal{D}(\hat\rho_o^{\rm tr},\hat\rho_o^{\rm ex}).
\end{align}

\bibliography{bbgky_lit}
\bibliographystyle{unsrt}

\end{document}